\documentclass[12pt]{article}
\usepackage{graphicx}
\usepackage{epsf,amsmath,bbold,amsfonts,stmaryrd}

\usepackage[utf8]{inputenc}
\usepackage{mathrsfs}
\usepackage{appendix}
\usepackage{amssymb}
\usepackage{float}
\usepackage{color}
\usepackage{cite}
\usepackage[colorlinks]{hyperref}
\hypersetup{pageanchor=false}
\usepackage{indentfirst}
\usepackage{url}
\usepackage{float}
\usepackage{caption}
\usepackage[numbers,square,comma,sort&compress,merge]{natbib}

\def\b{\beta}

\def\mc{\mathcal}

\def\t{\theta}

\def\be{\begin{equation}}
\def\ee{\end{equation}}

\def\bea{\begin{eqnarray}}
\def\eea{\end{eqnarray}}

\def\ba{\begin{array}}
\def\ea{\end{array}}

\def\bc{\begin{center}}
\def\ec{\end{center}}

\def\bs{\begin{split}}
\def\es{\end{split}}

\def\bl{\begin{flushleft}}
\def\el{\end{flushleft}}

\def\br{\begin{flushright}}
\def\er{\end{flushright}}

\def\bi{\begin{itemize}}
\def\ei{\end{itemize}}

\def\bt{\begin{tabular}}
\def\et{\end{tabular}}
\newcommand{\sR}{\mathsf{R}}
\newcommand{\sK}{\mathsf{K}}
\numberwithin{equation}{section}

\title{\textbf{Charged Black Rings at large $D$}}

\author{
Bin Chen$^{1,2,3}$\footnote{bchen01@pku.edu.cn},\,
Peng-Cheng Li$^{1}$\footnote{wlpch@pku.edu.cn}
and Zi-zhi Wang$^{1}$\footnote{1300011456@pku.edu.cn}
}
\date{}
\begin{document}
\maketitle

\begin{center}
{\it
$^{1}$Department of Physics and State Key Laboratory of Nuclear Physics and Technology,\\Peking University, 5 Yiheyuan Rd, Beijing 100871, P.R.\,China\\
\vspace{2mm}
$^{2}$Collaborative Innovation Center of Quantum Matter, 5 Yiheyuan Rd, \\Beijing 100871, P.R.\,China\\ \vspace{1mm}
$^{3}$Center for High Energy Physics, Peking University, 5 Yiheyuan Rd, \\Beijing 100871, P.R.\,China}
\vspace{10mm}
\end{center}

\vskip 1.2cm
\begin{abstract}
%\noindent
We study  the charged slowly rotating black holes in the Einstein-Maxwell theory in the  large  dimensions ($D$). By  using the $1/D$ expansion in the near region of the black hole we obtain the effective equations for the charged slowly rotating black holes. The effective equations describe the charged black ring, the charged slowly rotating Myers-Perry black hole and the charged slowly boosted black string as stationary solutions. By embedding the solution of the effective equations into the flat spacetime background  in the ring coordinate we obtain the charged black ring solution at large $D$ analytically. We find that the charge lowers the angular momentum of the black ring due to the regular condition on the solution. By the perturbation analysis of the effective equations,  we obtain the quasinormal modes of the charge perturbation and the gravitational perturbation analytically. Like the neutral case  the  charged thin black ring suffers from the Gregory-Laflamme-like instability under non-axisymmetric perturbations, but  the charge helps weaken the instability. Besides, we find that the large $D$ analysis always respect the cosmic censorship. 
\end{abstract}
\thispagestyle{empty}
\newpage

\section{Introduction}

The higher dimensional black holes have much more horizon topology and abundant physics \cite{ERblackholeinHD}. The uniqueness of the four-dimensional black hole is lost in higher dimensions. The protype is the five-dimensional black ring discovered by Emparan and Real  \cite{ERarotating}. The black ring is still asymptotically flat, but is of the horizon topology $S^1\times S^2$.  In the dimensions higher than five one naturally expect to find such black ring solution,  just like the case that the Myers-Perry (MP) black hole \cite{MP} as the higher-dimensional version of the Kerr solution. However, the exact black ring solution in the dimensions higher than five is still missing. Many efforts on the higher dimensional black ring solution come from the numerical analysis and the approximation  for thin black ring by using the blackfold method \cite{EHNOR}. The large $D$ (spacetime dimensions)  method  developed recently offers a promising approximation framework to address this issue.

 The basic idea underlying the large $D$ expansion\cite{EST:lageDlimit} is that when the spacetime dimension is very large  the radial gradient of the gravitational field around the black hole becomes dominant ($\partial_r\sim \frac{D}{r_0}$),  such that the gravitational field is localized within a short radial distance $\sim r_0/D$ ($r_0$ is the horizon size). This makes the perturbation problems such as the computation of the quasinormal modes (QNMs) \cite{Decoupleing, EST:QNMAdS, EST:rotating, CFLY:GB} much simpler. Moreover for the class of the decoupled perturbations \cite{Decoupleing} the black hole can be effectively regarded as a thin membrane \cite{EST:Effective, membrane} with thickness $\sim r_0/D$. The shape of the membrane that is the topology of the horizon is described by the embedding of the membrane into a background spacetime, and the non-linear dynamics of the membrane is determined by the effective equations obtained by integrating out the radial direction. By solving the effective equations with proper embeddings of the membrane, one can construct different black hole solutions and furthermore study  their dynamics  perturbatively to find the quasinormal modes or   determine the end points of their evolutions due to the instability numerically\cite{EST:Effective, membrane, chargedmembrane, Dandekar:2016fvw, StationarylargeD, BRlargeD,ElasticLargeD, EILT:Hydro, TanabeIndS, RE:OnBrane, Tanabe:chargedrotating, EST:evolution,CLInstaGB,Dandekar:2016jrp}.

 As a direct application of the large $D$ effective theory,  the author in \cite{BRlargeD} constructed a neutral black ring solution at large $D$ analytically by solving the effective equations for slowly rotating black hole (whose definition will be given below) and studied its dynamical instability. In this paper we would like to extend the work in \cite{BRlargeD} to the charged case in the Einstein-Maxwell theory at large $D$.  We have strong motivation to study the charged black ring at large $D$. First of all, though there have already been some studies on the charged black ring solutions  in five-dimensional supergravity theories (e.g.\cite{Elvang:chargedblackring, supersymmetricblackring}) and  numerical constructions (e.g \cite{KKS}) in  pure Einstein-Maxwell theory,  the exact form of a charged stationary black ring solution in pure Einstein-Maxwell theory is not known. Moreover,  the charged rotating black holes of a spherical horizon topology in higher dimension is poorly understood as well. For example, the slowly charged rotating black hole has been studied in \cite{Aliev}, in which the rotation is assumed to be slow and could be treated as a perturbation. Therefore, it would be interesting to discuss the charged rotating black objects in the large expansion.

Instead of specializing to a particular supergravity model, we shall consider the pure Einstein-Maxwell theory. Although this theory is generically non-supersymmetric, it enters all gauged supergravities as the basic building block. Therefore one can expect the basic features of its solutions to be generic. We would like to see whether it is possible to obtain a charged black ring or a charge black hole in the Einstein-Maxwell theory at large $D$. Moreover it would be interesting to see the physical implications of the charge on the solutions and their dynamical stabilities.

Roughly speaking, one can imagine a black ring as a rotating bent string such that the centrifugal force balances the string tension.   According to the analysis using the  blackfold method \cite{EHNOR}, a thin black ring can be identified to be a boosted black string and its  horizon velocity is determined to be
\be\label{angualrvecofthinring}
\Omega_H=\frac{1}{\sqrt{D-3}}\frac{1}{R},
\ee
where $R$ is a ring radius. The thin black ring means $R$ is much larger than the ring thickness. At large $D$ this is a universal result not just for the thin black ring, because in this case the string tension is small \cite{EST:lageDlimit} so is the horizon angular velocity of a large $D$ black ring.  Therefore, a large $D$  black ring has a horizon angular velocity of $\mc O(1/\sqrt{D})$. This estimation  works also for the charged black ring, since  the blackfold method  can be used to construct a thin charged black ring from a charged boosted black string \cite{CEV:Higherrot}  as well. Furthermore the charge can generate repulsive force to against the string tension. As a result, the horizon angular velocity in the charged case should be even smaller than (\ref{angualrvecofthinring}). The black holes with $\mc O(1/\sqrt{D})$ horizon angular velocity will be referred to as the slowly rotating  black holes.

In this paper, based on the assumption (\ref{angualrvecofthinring}), we will study the large $D$ effective theory of a class of the charged  slowly rotating black holes, including the charged black ring, the charged slowly single-rotating  black hole and the charged slowly boosted black string. In section \ref{EffEqs} we solve the  Einstein-Maxwell equations with proper metric and gauge field ansatz and obtain the effective equations for the charged slowly rotating large $D$ black holes. We obtain the stationary solutions of the effective equations and discuss thermodynamic quantities of the solutions. In section \ref{chargedblackring} we construct the charged black ring solution by solving the effective equations with the embedding of the leading order solution into flat spactime background in the ring coordinate. By perturbative analysis of the effective equations we obtain the QNMs of the charge perturbation and the gravitational perturbation around the charged black ring.  We find that the ring suffers from the Gregory-Laflamme-like instability but  the charge helps weaken the instability.  We have similar discussions  in section \ref{chargedalowlyBH}  on the charged slowly rotating  black hole and the charged slowly boosted black string respectively. Especially for the charged slowly boosted black string, we show it is related to the black ring in the large ring radius limit. We end with a summary and some discussions in section \ref{summary}.

\section{Effective equations}\label{EffEqs}

In this section we consider the large $D$ effective theory for charged slowly rotating black holes in the Einstein-Maxwell theory. By solving the Einstein-Maxwell equations we obtain the effective equations, which contains the information on the mass, the charge and the momentum density of a dynamical black object.  In the following for convenience we use $1/n$ as the expansion parameter instead of $1/D$, where
\be
n=D-4.
\ee

\subsection{Setup}\label{setup}
We would like to  solve the Einstein-Maxwell equations
\be\label{Einstein-Maxwelleq}
R_{\mu\nu}-\frac{1}{2}R^g g_{\mu\nu}=\frac{1}{2}\Big(F_{\mu\rho}F_\nu^{\,\rho} -\frac{1}{4}F_{\rho\sigma}F^{\rho\sigma}g_{\mu\nu}\Big),
\quad \nabla^\mu F_{\mu\nu}=0,
\ee
where $F_{\mu\nu}=\partial_\mu A_\nu-\partial_\nu A_\mu$. We make the following metric ansatz for the charged slowly rotating black holes in the ingoing Eddington-Finkelsteins coordinates
\be\label{metricansatz}
ds^2=-Adv^2+2(u_v dv+u_a dX^a)dr+r^2G_{ab} dX^adX^b-2C_a dv dX^a+r^2 H^2 d\Omega_n^2.
\ee
The gauge field ansatz is
\be\label{Maxwellansatz}
A_\mu dx^\mu=A_v dv+A_a dX^a.
\ee
Here $X^a$ denotes two coordinates $(z, \Phi)$ in which  $\Phi$ can be specified to be the rotational direction. As a gauge choice we can set  $A_r dr=0$ and $u_a dX^a dr=0$. Generally, all components of the metric and the gauge field are functions of $(v, r, X^a)$.

The above form of the metric ansatz is inspired by the form of the MP black hole with rotation parameter $a\sim \mc O(1/\sqrt{n})$.  Since we are considering  the slowly rotating black hole, even including the electric charge, the form of the metric keeps invariant.  This is similar to the case that up to leading order of $a$ the higher dimensional charged rotating black hole is easy to obtain\cite{Aliev}, because $a$ only appears in the component $g_{v\Phi}$ which can be  treated as a perturbation. Note that the information of the horizon shape in  (\ref{metricansatz}) is undetermined, which makes the existence of other solutions such as black ring and black string become possible.

In order to  make $1/n$ expansion of the Einstein-Maxwell equations properly we need to know the large $D$ scaling of each component of the metric and the gauge field. The subtlety concerns the order of $C_{\Phi}$. As we mentioned before, we would like to consider the slowly  rotating black holes, whose horizon velocity is of $\mc O(1/\sqrt{n})$, so we have
\be
C_{\Phi}\sim\mc O(1/\sqrt{n}).
\ee
 It is useful to introduce a new azimuthal coordinate $\phi$ defined by
\be
\phi=\sqrt{n}\, \Phi,
\ee
such that in the new coordinate
\be
g_{v\phi}=\mc O(1/n),
\ee
 which makes the $1/n$ expansion more convenient. In terms of $\phi$,  the metric and the gauge field ansatz can be rewritten as
\be
ds^2=-Adv^2+2u_v dvdr+r^2G_{ab}dx^adx^b-2C_advdx^a+r^2H^2d\Omega_n^2.
\ee
and
\be
A_\mu dx^\mu=A_v dv+A_a dx^a,
\ee
where $x^a=(z,\phi)$.   At large $D$ the radial gradient becomes dominant, that is $\partial_r=\mc O(n)$,
$\partial_v=\mc O(1)$, $\partial_a=\mc O(1)$, so in the near region of black hole it is better to use a new radial coordinate $\sR$ defined by
\be
\sR=\Big(\frac{r}{r_0}\Big)^n,
\ee
such that $\partial_\sR=\mc O(1)$, where $r_0$ is a horizon length scale which can be set to be unit $r_0=1$. The large $D$ scaling of  the metric and gauge field functions are respectively
\be
A= \mc O(1), \quad u_v=\mc O(1),\quad C_a=\mc O(n^{-1}), \quad G_{zz}=1+\mc O(n^{-1}),
\ee
\be
G_{z\phi}=\mc O(n^{-2}),\quad G_{\phi\phi}=\frac{G(z)^{2}}{n}\Big(1+\frac{G_{\phi\phi}^{(0)}}{n}+\mc O(n^{-2})\Big), \quad H=H(z),
\ee
and
\be
A_v= \mc O(1),\quad A_a= \mc O(n^{-1}).
\ee
Note that $G$ and $H$ only depend on $z$, which comes from the fact that at the asymptotic infinity  both $\partial_v$ and $\partial_\phi$ are the  Killing symmetries. Other metric functions and gauge field functions depend on $(v, \sR,x^a)$.

\subsection{Effective equations}\label{effectiveequations}

To the leading order  the Einstein-Maxwell equations (\ref{Einstein-Maxwelleq}) contain only $\sR$-derivatives, so we can integrate them easily to read the solutions.  After imposing the boundary conditions, the leading order solutions  are\footnote{In fact $G_{\phi\phi}^{(0)}$ is obtained at the next-to-leading order in the $1/n$ expansion of the Einstein-Maxwell equations. We list it here because it appears also at the leading order of the Einstein-Maxwell equations. }
\be\label{Leadingordersolution}
A=1-\frac{p_v}{\sR}+\frac{q^2}{\sR^2},\quad  A_v=\frac{\sqrt{2}q}{\sR},
\quad u_v=\frac{H(z)}{\sqrt{1-H'(z)^2}},
\ee
\be
 A_a=-\frac{1}{n}\frac{\sqrt{2}p_a\,q }{p_v}\frac{1}{\sR},\quad C_a=\frac{1}{n}\frac{p_a}{p_v}(\frac{p_v}{\sR}-\frac{q^2}{\sR^2}), \quad G_{zz}=1+\mc O(n^{-2}),
\ee
\be
G_{z\phi}=\frac{1}{n^2}\Bigg[\frac{p_z\,p_\phi}{p_v^2}(\frac{p_v}{\sR}-\frac{q^2}{\sR^2})
-u_v\Big(\frac{2G'}{G}\frac{p_\phi}{p_v}-\partial_z(\frac{p_\phi}{p_v})-\partial_\phi\big(\frac{p_z}{p_v}\big)\Big)
 \ln\Big(1-\frac{\sR_-}{\sR}\Big)\Bigg],
\ee
\be
\begin{split}
G_{\phi\phi}^{(0)}=\,&-\Big(2+\frac{2HG'H'}{G(1-H'^2)} \Big)\ln\sR+\frac{p_\phi^2}{G^2 p_{v}^{2}}\Big(\frac{p_v}{\sR}-\frac{q^2}{\sR^2}\Big)\\
& -\Big(2+\frac{2HG'H'}{G(1-H'^2)}+\frac{2H}{\sqrt{1-H'^2}G^2}\partial_\phi\big(\frac{p_\phi}{p_v}\big) \Big) \ln\Big(1-\frac{\sR_-}{\sR}\Big),
\end{split}
\ee
where
\be
\sR_-\equiv\frac{p_v-\sqrt{p_v^2-4q^2}}{2}.
\ee
is one of the real roots of $A=0$, with the other one being
\be
\sR_+\equiv\frac{p_v+\sqrt{p_v^2-4q^2}}{2}.
\ee
For the charged slowly rotating black holes  $\sR_+$ and $\sR_-$ can be regarded as the outer horizon and the inner horizon of the dynamical black hole respectively. The uncharged case corresponds to $q=\sR_-=0$, which reproduce the results in \cite{BRlargeD}.

In the above expressions, $p_v=p_v(v,x^a)$, $q=q(v,x^a)$ and $p_a=p_a(v,x^a)$ are the integration functions  of the leading order equations. Physically they  can be taken as  the mass, the charge and the momentum density.

Besides the four integration functions mentioned above, the functions $G(z)$ and $H(z)$ are undetermined. They actually encode the information of the topology of the event horizon. They can be determined by embedding the leading order solution into a specific background spacetime. As will be shown, the different embeddings decide different forms of the functions $G(z)$ and $H(z)$, and thus different black objects, either the black rings, or the black holes or the black strings.

By performing the ``$D-1+1$" decomposition on a $r=$constant surface, the momentum constraint tells us that the function $H(z)$ should satisfy the following relation
\be\label{momentumcons}
1-H'(z)^2+H(z)H''(z)=0.
\ee
As a direct result, $u_v$ can be treated as a constant, so we denote
\be\label{hatkappa}
\frac{\sqrt{1-H'(z)^2}}{H(z)}=2\hat{\kappa},
\ee
where $\hat{\kappa}$ is a constant and is related to the surface gravity of the horizon as we will see later.

At the next-to-leading order of $1/n$ expansion, we are able to find  non-trivial conditions on which $p_v$ and $p_a$ should satisfy. These conditions can be obtained equivalently from the momentum constraints on a $r$=constant surface  at the same order of $1/n$ expansion. The non-trivial conditions are just the  effective equations for the slowly rotating large $D$ black holes. These equations are listed as follows
\be\label{EffEq1}
\partial_vq+\frac{H'(z)}{H(z)}\frac{p_z q}{p_v}-\frac{H'(z)}{2\hat{\kappa}H(z)}\partial_zq-\frac{\partial_\phi^2q}{2\hat{\kappa}G(z)^2}+\frac{1}{G(z)^2}\partial_\phi\biggl[\frac{p_\phi q}{p_v}\biggl]=0,
\ee

\be\label{EffEq2}
\partial_v p_v+\frac{H'}{H}p_z -\frac{H'}{2\hat{\kappa}H}\partial_z p_v-\frac{\partial_\phi^2 p_v}{2\hat{\kappa}G(z)^2}+\frac{\partial_\phi p_\phi}{G(z)^2} =0,
\ee

\be\label{EffEq3}
\begin{split}
&\partial_v p_z-\frac{2p_z \sR_-}{p_v q}\partial_v q+\partial_z p_v -\frac{H'}{2\hat{\kappa}H}\partial_z p_z+\frac{H'\sR_-}{\hat{\kappa}Hp_v}\bigg[\partial_z p_z-\frac{p_z}{p_v}\partial_z p_v+\frac{p_z}{q}\partial_z q\biggl]+\frac{p_\phi}{G^2 p_v}\partial_\phi p_z\\
&\qquad+\biggl[\frac{p_\phi G'(3\sR_+-\sR_-)\sR_-}{\hat{\kappa}p_v^2G^3(\sR_+-\sR_-)}-\frac{p_\phi p_z(\sR_+-\sR_-)}{G^2p_v^3}\biggl]\partial_\phi p_v+\frac{(\hat{\kappa}Gp_z+p_vG')(\sR_+-\sR_-)}{\hat{\kappa}G^3p_v^2}\partial_\phi p_\phi\\
&\qquad-\bigg[\frac{2p_z p_\phi \sR_-}{G^2p_vq}+\frac{2p_\phi qG'}{\hat{\kappa}G^3p_v(\sR_+-\sR_-)}\biggl]\partial_\phi q-\frac{\sR_-}{2\hat{\kappa}G^2p_v^2}(\partial_z p_v )(\partial_\phi p_\phi)
-\frac{1}{2\hat{\kappa}G^2p_v }\partial_\phi^2 p_z\\
&\qquad+\frac{\sR_-}{2\hat{\kappa}G^2p_v}\Big[-\frac{p_z}{ p_v}\partial_\phi^2 p_v+\frac{p_z}{q}\partial_\phi^2q-\frac{p_\phi}{p_v}\partial_z\partial_\phi p_v +\partial_z\partial_\phi p_\phi+\partial_\phi^2 p_z\Big]\\
&\qquad+\frac{\sR_-^2}{2\hat{\kappa}G^2 p_v^2(\sR_+-\sR_-)}\Big[\frac{p_\phi}{p_v}(\partial_z p_v)(\partial_\phi p_v)
+\frac{p_z}{p_v}(\partial_\phi p_v)^2-(\partial_\phi p_v)(\partial_\phi p_z)\Big]\\
&\qquad+\frac{q}{\hat{\kappa}G^2p_v(\sR_+-\sR_-)}\Big[-\frac{p_z}{p_v}(\partial_\phi p_v)(\partial_\phi q)+(\partial_\phi p_z)(\partial_\phi q)
+(\partial_z p_\phi)(\partial_\phi q)\\
&\qquad-\frac{q}{p_v}(\partial_\phi p_v)(\partial_z p_\phi)-\frac{p_\phi}{p_v}(\partial_z p_v)(\partial_\phi q)\Big]
+\frac{(\sR_+-\sR_-)H'}{Hp_v^2}p_z^2-\frac{G'}{G^3p_v}p_\phi^2\\
&\qquad+\frac{2H'^2\sR_+-p_v}{2\hat{\kappa}H^2p_v}p_z+\frac{(\sR_+-\sR_-)\big(HG'^2H'+G(G'-HH'G'')\big)}{4\hat{\kappa}^2G^2H^2}=0,
\end{split}
\ee
and

\be\label{EffEq4}
\begin{split}
&\partial_v p_\phi-\frac{2p_\phi\sR_-}{p_v q}\partial_v q-\frac{p_\phi H'\sR_-}{2\hat{\kappa}Hp_v^2}\partial_z p_v-\frac{H'\sR_+}{2\hat{\kappa}Hp_v}\partial_z p_\phi+\frac{p_\phi H'\sR_-}{\hat{\kappa}Hp_v q}\partial_z q+\frac{H'\sR_-}{2\hat{\kappa}Hp_v}\partial_\phi p_z+\frac{2p_\phi\sR_+}{G^2p_v^2}\partial_\phi p_\phi\\
&-\Big[\frac{p_\phi^2(\sR_+-\sR_-)}{G^2p_v^3}+\frac{p_v}{(\sR_+-\sR_-)}+\frac{(\hat{\kappa}Gp_z+p_vG')H'}{4\hat{\kappa}^2GHp_v}
-\frac{p_z(\sR_+-\sR_-)H'}{4\hat{\kappa}Hp_v^2}+\frac{p_vG'H'}{4\hat{\kappa}^2GH(\sR_+-\sR_-)}\Big]\partial_\phi p_v\\
&
+\Big[-\frac{2p_\phi^2\sR_-}{G^2p_v^2q}+\frac{4q}{(\sR_+-\sR_-)}+\frac{qG'H'}{\hat{\kappa}^2GH(\sR_+-\sR_-)}\Big]\partial_\phi q
-\frac{(\sR_+-\sR_-)}{2\hat{\kappa}G^2p_v}\partial_\phi^2 p_\phi-\frac{p_\phi\sR_-}{\hat{\kappa}G^2p_v^2}\partial_\phi^2 p_v\\
&+\frac{p_\phi\sR_-}{\hat{\kappa}G^2p_vq}\partial_\phi^2 q+\frac{\sR_{-}^2}{\hat{\kappa}G^2p_v^2(\sR_+-\sR_-)}\Big[\frac{p_\phi}{p_v}(\partial_\phi p_v )^2
-(\partial_\phi p_v)(\partial_\phi p_\phi)\Big]\\&-\frac{2 q}{\hat{\kappa}G^2p_v(\sR_+-\sR_-)}\Big[\frac{p_\phi}{p_v}(\partial_\phi p_v)(\partial_\phi q)-(\partial_\phi p_\phi)(\partial_\phi q)\Big]\\&
+\Big[\frac{G'H'}{2\hat{\kappa}GH}
+\frac{(\sR_+-\sR_-)(2\hat{\kappa}Gp_z+p_vG')H'}{2\hat{\kappa}GHp_v^2}\Big]p_\phi=0.
\end{split}
\ee
It is easy to see that if we set $q=0$, so $\sR_+=p_v$ and $\sR_-=0$, then we obtain the effective equations of the uncharged case found in \cite{BRlargeD}. From these equations  we can obtain the stationary solutions and the  non-linear evolution of the dynamical black holes. But as we mentioned before we can not completely solve these equations without knowing the form of $G(z)$ and $H(z)$.

\subsection{Stationary solutions}
To obtain the stationary solutions we need to assume  the existence of two Killing vectors $\sK_v=\partial_v$ and $\sK_\phi=\partial_\phi$, then we have
\be
p_v=p_v(z),\quad q=q(z),\quad p_a=p_a(z).
\ee
Now the above effective equations are simplified significantly to the following forms if $H'\neq 0$
\be\label{EffEq1r}
\frac{p_z q}{p_v}-\frac{1}{2\hat{\kappa}}\partial_z q=0,
\ee

\be\label{EffEq2r}
p_z =\frac{1}{2\hat{\kappa}}\partial_z p_v
\ee

\be\label{EffEq3r}
\begin{split}
&\partial_z p_v -\frac{H'}{2\hat{\kappa}H}\partial_z p_z+\frac{H'\sR_-}{\hat{\kappa}Hp_v}\bigg[\partial_z p_z-\frac{p_z}{p_v}\partial_z p_v+\frac{p_z}{q}\partial_z q\biggl]
+\frac{(\sR_+-\sR_-)H'}{Hp_v^2}p_z^2-\frac{G'}{G^3p_v}p_\phi^2\\
&\qquad+\frac{2H'^2\sR_+-p_v}{2\hat{\kappa}H^2p_v}p_z+\frac{(\sR_+-\sR_-)\big(HG'^2H'+G(G'-HH'G'')\big)}{4\hat{\kappa}^2G^2H^2}=0,
\end{split}
\ee
and
\be\label{EffEq4r}
\begin{split}
&-\frac{p_\phi \sR_-}{p_v^2}\partial_z p_v-\frac{\sR_+}{p_v}\partial_z p_\phi+\frac{2p_\phi \sR_-}{p_v q}\partial_z q
+\Big[\frac{G'}{G}
+\frac{(\sR_+-\sR_-)(2\hat{\kappa}Gp_z+p_vG')}{Gp_v^2}\Big]p_\phi=0.
\end{split}
\ee
From (\ref{EffEq1r}) and (\ref{EffEq2r}) we have
\be\label{Stationarysolutionq}
 q(z)=Q p_v(z),
 \ee
 where $Q$ is an integration constant.
Plugging (\ref{EffEq2r}) and (\ref{Stationarysolutionq}) into (\ref{EffEq4r}) we obtain
 \be\label{Stationarysolutionpphi}
 p_\phi(z)=\hat{\Omega}_HG(z)^2p_v(z).
 \ee
 Here $\hat{\Omega}_H$ is a constant independent of $z$, whose physical meaning will become  manifest from the metric. Let us write down the $(v, \Phi)$ part of the leading order metric of the above stationary solutions
 \be
 ds^2_{(v,\Phi)}=-Adv^2+\Big(d\Phi-\frac{\hat{\Omega}_H}{\sqrt{n}}(1-A)dv \Big)^2,
 \ee
 where $A$ is given by (\ref{Leadingordersolution}). It is easy to see that
 \be
 \Omega_H=\hat{\Omega}_H/\sqrt{n},
 \ee
  which is exactly the same as the uncharged case shown in \cite{BRlargeD}. Then we can see that the event horizon at the surface
   $\sR=\sR_+$, is a Killing horizon of the Killing vector
 \be
 \xi=\partial_v+\hat{\Omega}_H\partial_\phi.
 \ee
 Due to the existence of the Killing vector $\xi$, the associated surface gravity is given by
 \be\label{surfacegra}
 \begin{split}
 \kappa&=-\frac{\partial_r(\xi_\mu \xi^\mu)}{2\xi_r}\Big|_{\sR=\sR_+}\\
&=n\hat{\kappa}\frac{\sqrt{1-4Q^2}+4Q^2-1}{2Q^2}\biggl[1+\mc O(1/n)\biggl],
 \end{split}
 \ee
 which is proportional to $\hat{\kappa}$.
 As $\hat{\kappa}$ is a constant, so is the surface gravity, which is expected for the stationary solutions. Now the function $A$ is of the form
 \be
 A=1-\frac{p_v}{\sR}+\frac{Q^2p^2_v}{\sR^2},
 \ee
 whose roots are
 \be
 \sR_\pm=\frac{1\pm \sqrt{1-4Q^2}}{2}p_v.
 \ee
 In order to have physical solutions without naked singularity, we must require
 \be\label{chargecondition}
 1-4Q^2\geq 0,
 \ee
 with the equality corresponding to the extremal solutions. This is very similar to the charged black hole case, as now the rotating is very slowly and can be treated as a perturbation. Obviously, the parameter $p_v$ could be related to the mass. It can be determined from the last effective equation (\ref{EffEq3r}). By setting $p_v(z)=e^{P(z)}$, we obtain an ordinary differential equation
  \be
  \begin{split}
  &P''(z)-\frac{H'(z)}{H(z)}P'(z)\\
  &-\biggl[\frac{G'(z)^2}{G(z)^2}+\frac{G'(z)}{G(z)H(z)H'(z)}-\frac{G''(z)}{G(z)}-\hat{\Omega}_H^2\frac{G(z)G'(z)(1-H'(z)^2)}{\sqrt{1-4Q^2}H(z)H'(z)}\biggl]=0.
  \end{split}
  \ee
   if $1-4Q^2\neq0$. In order to solve this equation we need the exact form of $H(z)$ and $G(z)$. However, if $1-4Q^2=0$, which corresponds to the extremal limit, the above equation is not valid anymore. Instead,  (\ref{EffEq3r}) is reduced to
  \be\label{Effextremal}
  \hat{\Omega}_H^2\,G(z)G'(z)p_v=0.
  \ee
It leads to either the trivial solution $p_v=0$, or the vanishing angular velocity $ \hat{\Omega}_H=0$, or  the special case $G=$constant. In the latter two cases, there is no constraint on the function $p_v(z)$.   In the static case $ \hat{\Omega}_H=0$, there is a uniqueness theorem  on the charged solution in higher dimensional Einstein-Maxwell theory, the single black hole should be of spherical topology\cite{Rogatko:2003kj,Hollands:2012xy}. Therefore the extreme case is singular and one should take the extreme limit carefully.
%This situation indicates that the large $D$ method cannot  be directly applied to the extremal case. All we can do is  to extrapolate the result of  the non-extremal case to the extremal limit and see if it make sense. %At least we know the extrapolation is valid for the static case.

For $\hat{\Omega}_H=0$, the first two stationary equations (\ref{EffEq1r}) and (\ref{EffEq2r}) keep invariant, the fourth one  (\ref{EffEq4r}) vanishes identically, and the third one  (\ref{EffEq3r}) becomes 
 \be
\sqrt{1-4Q^2}\Bigg[P''(z)-\frac{H'(z)}{H(z)}P'(z)-\Big(\frac{G'(z)^2}{G(z)^2}+\frac{G'(z)}{G(z)H(z)H'(z)}-\frac{G''(z)}{G(z)}\Big)\Bigg]=0,
  \ee
after setting $p_v(z)=e^{P(z)}$. Both the extremal and non-extremal cases are described by this equation. For the extremal case $1-4Q^2=0$, the above equation is satisfied identically. However, for the non-extremal case, even the near extremal one $1-4Q^2\ll1$, the above equation is reduced to
  \be
P''(z)-\frac{H'(z)}{H(z)}P'(z)-\Big(\frac{G'(z)^2}{G(z)^2}+\frac{G'(z)}{G(z)H(z)H'(z)}-\frac{G''(z)}{G(z)}\Big)=0.
  \ee
Although the large $D$ method cannot  be directly applied to the extremal case, it is reasonable to extrapolate the result of  the non-extremal case to the extremal limit. However, one should take the limit carefully, as we will show later. % at least the well known RN black hole and charged black string solution show this point.

  From the above general stationary solutions we can also read the corresponding thermodynamic quantities by using the Komar integral approach. First let us write down the leading order metric
  \be\label{LOmetric}
   \begin{split}
  ds^2=&-\Big(1-\frac{p_v}{\sR}+\frac{Q^2p_v^2}{\sR^2}\Big)dv^2+\frac{dvdr}{\hat{\kappa}}+r^2G(z)^2d\Phi^2-2\Omega_HG(z)^2\Big(\frac{p_v}{\sR}-\frac{Q^2p_v^2}{\sR^2}\Big)dvd\Phi\\
  &+r^2dz^2+ r^2H^2d\Omega_n^2,
   \end{split}
  \ee
  and the leading order gauge field
  \be\label{gaugefiledLO}
  A_\mu dx^\mu=\frac{\sqrt{2}Qp_v}{\sR}dv-\frac{\sqrt{2}\Omega_HG(z)^2p_v}{\sR}d\Phi.
  \ee
  Here we omit $\mc O(1/n)$ terms. The horizon area  $\mathcal{A}_H$ is given by
\be\label{horizonarea}
\mathcal{A}_H=\pi(1+\sqrt{1-4Q^2})\Omega_n\int dz\,G(z)H(z)^np_v(z).
\ee
The mass and the angular momentum are given respectively by
\bea\label{mass}
\mathcal{M}&=&\frac{n\hat{\kappa}\Omega_n}{4\mathrm{G}_D}\int dz\, G(z)H(z)^n p_v(z),\\
\mathcal{J}&=&\frac{n\hat{\kappa}\Omega_n}{4\mathrm{G}_D}\int dz\,\Omega_H G(z)^3H(z)^n p_v(z).\label{angularmomentum}
\eea
From the expression of the angular momentum, it seems that the angular  momentum and the charge are independent of each other. As we shall see later, this is not exactly correct for the charged black ring, in which they are related to each other by the determination of the horizon angular velocity $\Omega_H $. The charge can be also easily obtained from (\ref{gaugefiledLO}) as
\be\label{blackholecharge}
\mathcal{Q}=\frac{\sqrt{2}n\hat{\kappa}\Omega_n}{4\mathrm{G}_D}\int dz\, QG(z)H(z)^n p_v(z),
\ee
and the electric potential on the horizon is given by
\be\label{electricpotential}
\Phi_H=A_\mu \xi^\mu\Big|_{\sR=\sR_+}=\frac{\sqrt{2}(1+\sqrt{1-4Q^2})}{2Q}\Big(1-\Omega_HG(z)^2\Big).
\ee
From these results we see that thermodynamic quantities of the charged slowly rotating black holes satisfy the Smarr formula at the leading order in $1/n$ expansion as
\be
\mathcal{M}=\frac{\kappa}{8\pi \mathrm{G}_D}\mathcal{A}_H+\Omega_H \mathcal{J}+\Phi_H \mathcal{Q}.
\ee
As the angular velocity $\Omega_H$ is of $\mc O(1/\sqrt{n})$, so $\Omega_H \mathcal{J}$ is of  $\mc O(1)$ and does not contribute to the Smarr formula at the leading order.

 \section{Charged black ring}\label{chargedblackring}

 In this section, we study the charged black ring at large $D$. We first construct the black ring solution by finding the functions  $G(z)$ and $H(z)$. Then we discuss the stability of the
 solution and show that the presence of charge may weaken the instability.

 \subsection{Charged black ring solution}\label{chargedblackringsolution}

 The $D=n+4$ dimensional flat metric in the ring coordinate \cite{blackring} is of the form
 \be\label{ringcoordinate}
 ds^2=-dt^2+\frac{R^2}{(R+r \mathrm{cos}\t)^2}\biggl[\frac{R^2dr^2}{R^2-r^2}+(R^2-r^2)d\Phi^2+r^2(d\t^2+\mathrm{sin}^2\t d\Omega_n^2)\biggl],
 \ee
 where $R$ is the ring radius, $0\leq r\leq R$, $0\leq\t\leq\pi$ and $0\leq \Phi\leq2\pi$. $r=0$ is the origin of the ring coordinate.
 On the other hand, in the far region of the black hole where $\sR\gg1$, the leading order metric we obtained before behaves like
 \be\label{leadingaymp}
 ds^2\simeq -dv^2+\frac{2H}{\sqrt{1-H'^2}}dvdr+r^2dz^2+r^2G^2d\Phi^2+r^2 H^2 d\Omega^2_n.
 \ee
On a $r=$ constant surface in the far region the induced metric from asymptotic  spacetime background  and the one from the near region solution in the large $D$ expansion should  match. Considering the surface $r=r_0=1$, which is in the far region since  $\sR\gg1$ still corresponds to $r=r_0+\mc O(1/n)$, and comparing (\ref{ringcoordinate}) with (\ref{leadingaymp}),  we  obtain the following identifications
 \be\label{identification}
 H=\frac{R\sin\t}{R+\cos\t},\quad G=\frac{R\sqrt{R^2-1}}{R+\mathrm{cos}\t},\quad \frac{d\t}{dz}=\frac{R+\mathrm{cos}\t}{R}.
 \ee
Hence $G(z)$ and $H(z)$ are determined. From  (\ref{hatkappa}) it is easy to obtain
 \be
 \hat{\kappa}=\frac{\sqrt{R^2-1}}{2R}.
 \ee
 For a stationary solution the associated surface gravity is still given by (\ref{surfacegra}).
 By these identifications we can obtain the leading order charged black ring solution as long as we solve the effective equations. Since $R\ge r$,  $R\ge 1$. The case that $R\gg1$ corresponds to thin black ring, while the case that  $R\simeq1$ corresponds to fat black ring.

 To solve the effective equations it is convenient to introduce the variable
\be
y=\cos\,\t.
\ee
 For the stationary solution with $\partial_v$ and $\partial_\phi$ being the Killing vectors,  we have
 \be
 p_v=e^{P(y)},\quad p_z=p_z(y),\quad p_\phi=p_\phi(y),\quad q=q(y).
 \ee
 From (\ref{EffEq2r}) we find
 \be\label{blackringpz}
 p_z(y)=-\frac{(R+y)\sqrt{1-y^2}}{\sqrt{R^2-1}}p_v'(y).
 \ee
 From (\ref{EffEq1r}) we have
 \be\label{blackringq}
 q(y)=Qp_v(y).
 \ee
 And then from (\ref{EffEq4r}) we obtain
 \be
 p_\phi(y)=\hat{\Omega}_H\frac{R^2(R^2-1)}{(R+y)^2}p_v(y).
 \ee
 Furthermore (\ref{EffEq3r}) gives an equation for $P(y)$ as
 \be
  \sqrt{1-4Q^2}\Big(-R+2(1+Ry)P'(y)+(R+y)(1+Ry)P''(y)\Big)+\frac{R^2(R^2-1)^2\hat{\Omega}_H^2}{(R+y)^3}=0.
 \ee
 If $1-4Q^2\neq0$, we have
 \be\label{Py}
 P''(y)+\frac{2}{R+y}P'(y)-\frac{R}{(R+y)(1+R y)}+\hat{\Omega}_H^2\frac{R^2(R^2-1)^2}{\sqrt{1-4Q^2}(R+y)^4(1+Ry)}=0.
 \ee
The above equation contains a pole at $y=-1/R$, which leads to a logarithmic divergence at $y=-1/R$ for the solution $P(y)$. In order to have a regular solution we need $\hat{\Omega}_H$ to be
 \be\label{balancecondition}
 \hat{\Omega}_H=(1-4Q^2)^{1/4}\frac{\sqrt{R^2-1}}{R^2}.
 \ee
In fact in some supergravity settings, for the five dimensional charged black rings \cite{Elvang:chargedblackring, supersymmetricblackring} and the higher dimensional thin charged black rings constructed by the blackfold method\cite{EHNOR, blackring}, the similar identity has been derived from the regularity of solution and dynamical balance condition of the boosted black string. From this expression  and (\ref{angularmomentum}) we can read that, with the increase of the charge $Q$, the angular momentum needed to balance the tension of the black ring decreases. This is consistent with the intuitive picture that the electric repulsive force can be used to contend with the attractive tension of the rotating black string.

Under  the above condition the solution to the equation (\ref{Py}) is given as
 \be\label{Pysol}
 P(y)=P_0+\frac{P_1}{R+y}+\frac{(1+Ry)(1+Ry+2R(R+y)\ln(R+y))}{2R^2(R+y)^2},
 \ee
where $P_0$ and $P_1$ are the integration constants. $P_0$ is the redefinition of $r_0$ and $P_1$ comes from the redefinition of the $\phi$ coordinate of the ring coordinate \cite{BRlargeD}. Summarizing the above results,  the metric of  the charged black ring solution at the leading order in the large $D$ expansion is of the following form
\be
\begin{split}
ds^2=&-\Big(1-\frac{p_v(z)}{\sR}+\frac{Q^2p_v(z)^2}{\sR^2}\Big)dv^2+\frac{dvdr}{\hat{\kappa}}\\
&+2\Big[\frac{(R+y)\sqrt{1-y^2}p_v'(y)}{\sqrt{R^2-1}}\Big(\frac{1}{\sR} -\frac{Q^2 p_v(y)}{\sR^2}\Big)\frac{dz}{n}-\frac{\hat{\Omega}_HR^2(R^2-1)}{(R+y)^2}\Big(\frac{p_v(y)}{\sR} -\frac{Q^2 p_v(y)^2}{\sR^2}\Big)\frac{d\phi}{n}\Big]dv\\
&-r^2\frac{2\hat{\Omega}_H\sqrt{1-y^2}\sqrt{R^2-1}R^2p_v'(y)}{R+y}\Big(\frac{1}{\sR} -\frac{Q^2 p_v(y)}{\sR^2}\Big)\frac{dzd\phi}{n^2}\\
&+r^2\frac{R^2(R^2-1)}{(R+y)^2}\Big[1-\frac{2R(R+y)}{R^2-1}\frac{\ln \sR}{n}+\hat{\Omega}_H^2\frac{R^2(R^2-1)}{n(R+y)^2}\Big(\frac{p_v(y)}{\sR}-\frac{Q^2 p_v(y)^2}{\sR^2}\Big)\\
&\qquad-\frac{2R(R+y)}{n(R^2-1)}\ln\Big(1-\frac{\sR_-}{\sR}\Big) \Big]\frac{d\phi^2}{n}+r^2 dz^2+\frac{r^2R^2(1-y^2)}{(R+y)^2}d\Omega_n^2,\label{ringsolution}
\end{split}
\ee
and the gauge field is of the form
\be\label{ringsolutiongaugefield}
A_\mu dx^\mu=\frac{\sqrt{2}Qp_v(y)}{\sR}dv+\frac{1}{n}\frac{(R+y)\sqrt{1-y^2}}{\sqrt{R^2-1}}\frac{\sqrt{2}Qp_v'(y)}{\sR}dz
-\frac{1}{n}\frac{\hat{\Omega}_HR^2(R^2-1)}{(R+y)^2}\frac{\sqrt{2}Qp_v(y)}{\sR}d\phi.
\ee
Similar to the uncharged case \cite{BRlargeD}, the solution blows up at $R=1$, which indicates that the large $D$ expansion is not valid for the very fat black ring ($R\to1$). But  other cases  can be described well by the solution, including the thin
black ring ($R\gg1$) and the  black ring not very thin ($R>1$).

If $1-4Q^2=0$, that is the extremal limit, (\ref{Py}) is not valid anymore, instead  there is condition
\be
\hat{\Omega}_H=0.
 \ee
Now the solution becomes static. Physically, this is reasonable as we can treating the rotation as a small perturbation in the charged black hole background, so in the extreme charge limit, the rotation should be vanishing, otherwise it would lead to the violation of cosmic censorship. In the $\hat{\Omega}_H \to 0$ limit, the solution  (\ref{ringsolution}) is still well-defined but it has a {\it null singularity}  at the event horizon. The similar phenomenon also occurs  for the  boosted black string and in \cite{Elvang:chargedblackring}. We may think that the singular behavior is a property of the solution itself instead of being brought by the limitation of the large $D$ method.

\subsection{Phase diagram}

From the charged black ring solution we obtained by the large $D$ expansion, we are able to draw the phase diagram of the solution. The thermodynamic quantities for a stationary solution have already been given  in section \ref{effectiveequations}.  Using the
 embedding (\ref{identification}) and the relation  \be
 dz=-\frac{R}{R+y}\frac{1}{\sqrt{1-y^2}}dy,
 \ee
we can obtain the corresponding thermodynamic quantities for the charged black ring.
The horizon area is given by
 \be
 \mathcal{A}_H=2\pi \Omega_n \hat{\mathcal{A}}_H,
 \ee
 where
 \be\label{hatAH}
\hat{\mathcal{A}}_H=\frac{1+\sqrt{1-4Q^2}}{2}\int dy \frac{R^2\sqrt{R^2-1}e^{P(y)}}{(R+y)^2\sqrt{1-y^2}}\Bigg(\frac{R\sqrt{1-y^2}}{R+y}\Bigg)^n.
 \ee
 The mass, the angular momentum and the charge are given respectively by
 \be
 \mathcal{M}=\frac{n\Omega_n}{8 \mathrm{G}_D}\hat{\mathcal{M}},\quad \mathcal{J}=\frac{\sqrt{n}\Omega_n}{8 \mathrm{G}_D}\hat{\mathcal{J}},
 \quad   \mathcal{Q}=\frac{\sqrt{2}n\Omega_n Q}{8 \mathrm{G}_D}\hat{\mathcal{M}},
 \ee
 where
 \be\label{hatM}
 \hat{\mathcal{M}}=\int dy \frac{R(R^2-1)e^{P(y)}}{(R+y)^2\sqrt{1-y^2}}\Bigg(\frac{R\sqrt{1-y^2}}{R+y}\Bigg)^n,
 \ee
 and
 \be\label{hatJ}
 \hat{\mathcal{J}}=\int dy \frac{(1-4Q^2)^{1/4}R(R^2-1)^{5/2}e^{P(y)}}{(R+y)^4\sqrt{1-y^2}}\Bigg(\frac{R\sqrt{1-y^2}}{R+y}\Bigg)^n.
 \ee
 From (\ref{surfacegra}), the temperature of the charged black ring is given by
 \be
 T=\frac{\kappa}{2\pi}=\frac{n}{4\pi}\frac{\sqrt{R^2-1}}{R}\frac{\sqrt{1-4Q^2}+4Q^2-1}{2Q^2}.
 \ee
 The horizon angular velocity of the charged black ring is given by
 \be
 \Omega_H=\frac{\hat{\Omega}_H}{\sqrt{n}}=\frac{(1-4Q^2)^{1/4}}{\sqrt{n}}\frac{\sqrt{R^2-1}}{R^2}.
 \ee
 According to \cite{EHNOR}, we can define  the following dimensionless quantities: the spin $j$, the area $a_H$, the angular velocity $\omega_H$ and the temperature $t_H$, by fixing the mass as a basic scale  such that the charge is also fixed for a given charge parameter $Q$
 \be
 j^{n+1}=c_j\frac{\mathcal{J}^{n+1}}{\mathrm{G}_D\mathcal{M}^{n+2}},\qquad a_H^{n+1}=c_a\frac{\mathcal{A}^{n+1}_H}{(\mathrm{G}_D\mathcal{M})^{n+2}}
 \ee
 \be
 \omega_H=c_\omega \Omega_H (\mathrm{G}_D\mathcal{M})^{\frac{1}{n+1}},\qquad t_H=c_t(\mathrm{G}_D\mathcal{M})^{\frac{1}{n+1}} T,
 \ee
 where the numerical constants  are defined by
 \be
 c_j=\frac{\Omega_{n+1}}{2^{n+5}}\frac{(n+2)^{n+2}}{(n+1)^{(n+1)/2}},\qquad
 c_a=\frac{\Omega_{n+1}}{2(16\pi)^{n+1}}\frac{n^{(n+1)/2}(n+2)^{n+2}}{(n+1)^{(n+1)/2}},
 \ee
 and
 \be
 c_\omega=\sqrt{n+1}\Big(\frac{n+2}{16}\Omega_{n+1}\Big)^{-\frac{1}{n+1}},\qquad
 c_t=4\pi\sqrt{\frac{n+1}{n}}\Big(\frac{n+2}{2}\Omega_{n+1}\Big)^{-\frac{1}{n+1}}.
 \ee
 \begin{figure}[t]
 \begin{center}
  \includegraphics[width=65mm,angle=0]{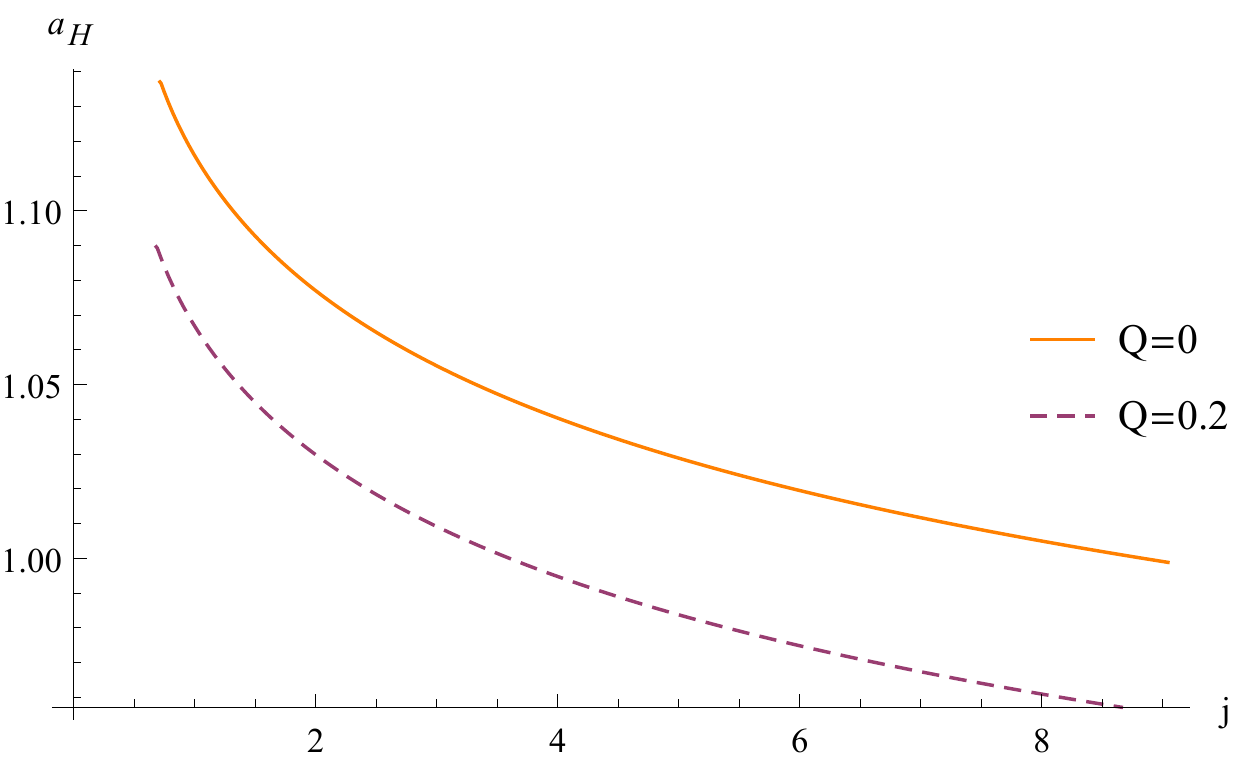}
 \end{center}
 \vspace{-5mm}
 \caption{The $(j, a_H)$ phase diagram of the large $D$ charged black ring solution for $n=20$($D=24$). The ring radius $R$ runs from $1.1$ to $20$, so both  the not-thin region and thin region ($R\gg1$) are covered. The solid curve shows the behavior of the uncharged case $Q=0$. The dashed line shows the behavior of the charged case with $Q=0.2$. }
 \label{figjaH}
\end{figure}
 Then we can easily evaluate $j$, $a_H$, $t_H$ and $\omega_H$ numerically for the large $D$ charged black ring solution by using
 (\ref{hatAH}), (\ref{hatJ}) and (\ref{hatM}), where $P(y)$ is given by (\ref{Pysol}). For convenience we  set $P_0=P_1=0$.
 Figure \ref{figjaH} shows the phase diagram of $(j, a_H)$ for the charged black ring solution with different charge parameters, where $R$ runs from $1.1$ to $20$ and $n=20$. The solid curve corresponds to the uncharged case $Q=0$, and the dashed curve corresponds to
 $Q=0.2$. Qualitatively two curves share the same  behavior. But the effect of the charge is also clear: both $a_H$ and $j$ decrease with the charge.

  \begin{figure}[t]
 \begin{center}
  \includegraphics[width=65mm,angle=0]{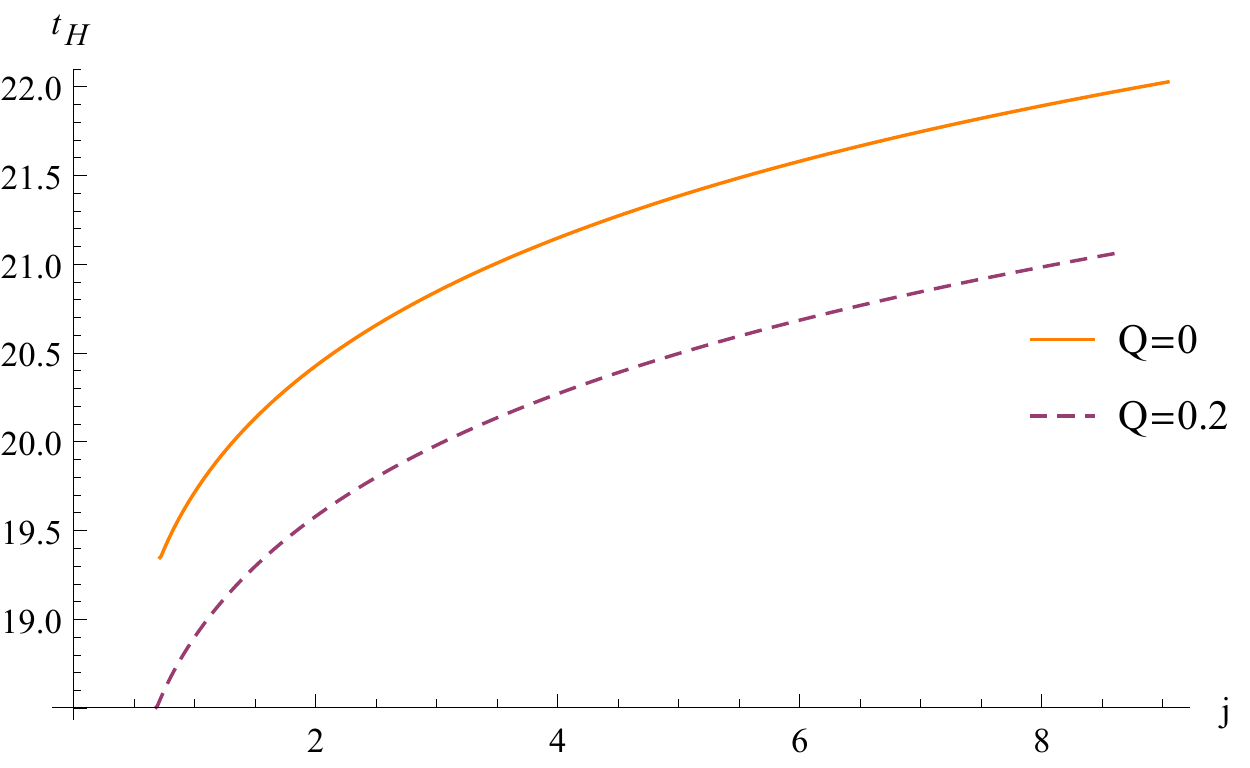}
  \hspace{5mm}
  \includegraphics[width=65mm,angle=0]{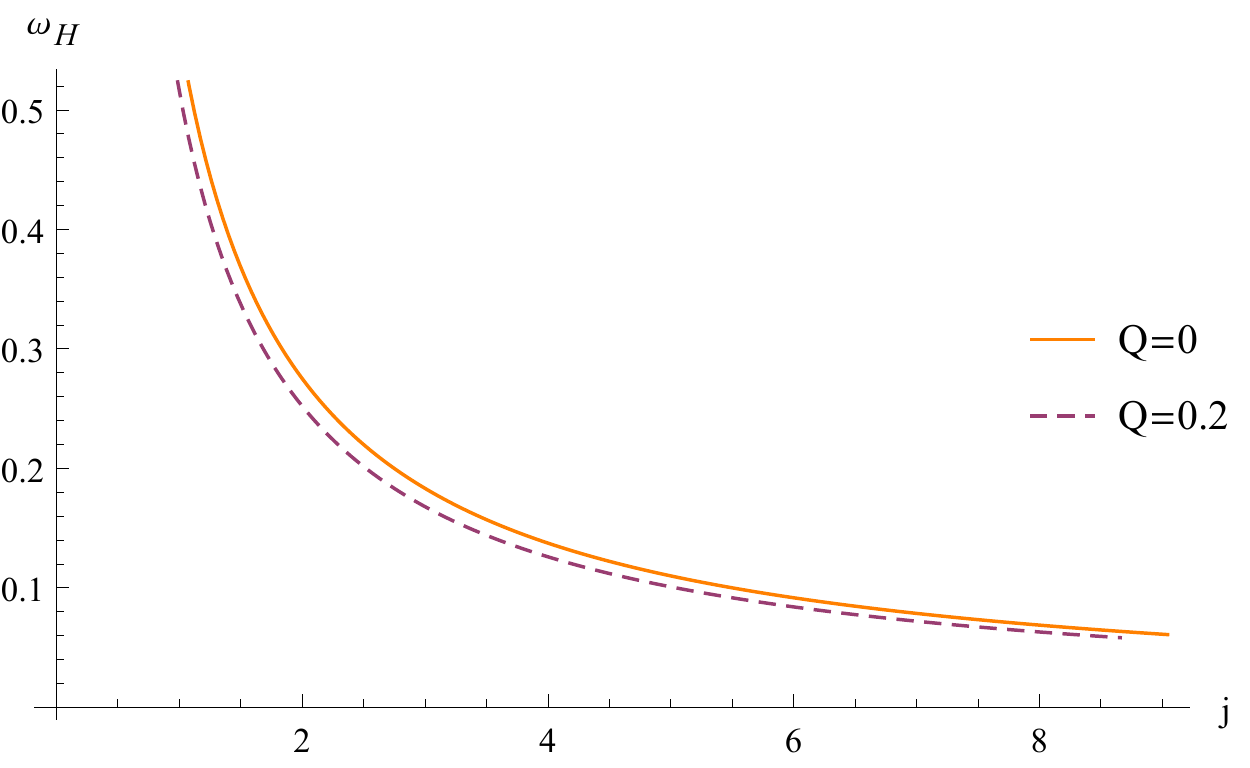}
 \end{center}
 \vspace{-5mm}
 \caption{The $(j, t_H)$ (left panel)and $(j, \omega_H)$ (right panel) phase diagram of the large $D$ black ring solution.
 The solid  and dashed line corresponds to the uncharged (Q=0) and charged case (Q=0.2) respectively. }
 \label{figjtH}
\end{figure}
Figure \ref{figjtH} shows the phase diagram of $(j, t_H)$ and $(j, \omega_H)$ for the large $D$ charged black ring solution. It is clear that the charged chase with $Q=0.2$ is similar to the uncharged case, and the charge lowers both $t_H$ and $\omega_H$.

\subsection{Quasinormal modes}

Taking a further step, now we would like to study the quasinormal modes of the charged black ring solution we obtained before. The quasinormal modes are obtained by the perturbation analysis
of the effective equations around the stationary charged black ring solution. The perturbation ansatz is
\bea
q(v,y,\phi)&=&Qe^{P(y)}\Big(1+\epsilon e^{-i\omega v} e^{im\phi} F_q(y)\Big),\\
p_v(v,y,\phi)&=&e^{P(y)}\Big(1+\epsilon e^{-i\omega v} e^{im\phi} F_v(y)\Big),\\
p_z(v,y,\phi)&=&-\frac{(R+y)\sqrt{1-y^2}}{\sqrt{R^2-1}}e^{P(y)}P'(y)\Big(1+\epsilon e^{-i\omega v} e^{im\phi} F_z(y)\Big),\\
p_\phi(v,y,\phi)&=& (1-4Q^2)^{1/4} \frac{(R^2-1)^{3/2}}{(R+y)^2}e^{P(y)}\Big(1+\epsilon e^{-i\omega v} e^{im\phi} F_\phi(y)\Big).
\eea
There is a subtlety about the order of the azimuthal quantum number $m$ associated with $\partial_\phi$. The original coordinate in the metric is $\Phi$, whose range is $0\leq\Phi\leq 2\pi$, hence the associated azimuthal number $m_\Phi$ is quantized as an integer. Due to the relation of $\phi$ and $\Phi$,  $m_\Phi$ is related with $m$ by
\be
m_\Phi=\sqrt{n} m.
\ee
Therefore $m$ does not have to be an integer unless $m_\Phi$ is as large as $\mc O(\sqrt{n})$. In \cite{BRlargeD}, the author considered the case with $m=\mc O(1)$, that is $m_\Phi=\mc O(\sqrt{n})$, it turns out that the only unstable mode is the GL-like instability. However the case $m_\Phi=\mc O(1)$ is also important since now a new elastic-type (named by the property that this mode deforms the black ring without substantially changing its thickness) instability emerges, as first found by the numerical study in \cite{FKT}, and then by the large $D$ study in \cite{ElasticLargeD} . To investigate the elastic instability, the $\mc O(1/n^2)$ corrections to the
effective equations are required. This is more complicated for the charged case, so in this paper we only consider the normal GL-like instability and wish that we may come back to study the more involved new elastic-type instability in the future.

Since we restrict ourself to the case with $m=\mc O(1)$, the form for the perturbations around the stationary solutions is appropriate. Substituting the above perturbations into the effective equations, and keeping the leading order of $\epsilon$, we obtain four ordinary differential equations for $F_v(y)$, $F_q(y)$, $F_z(y)$ and $F_\phi(y)$ with $\omega$ being undetermined. In order to solve the perturbation equations we need to impose the boundary conditions. Due to the singular behavior of the effective equations at $y=-1/R$, we could  assume that at the pole
 $y=-1/R$, $F_v(y)$ behaves like
 \be\label{blackringBC}
 F_v(y)\propto(1+Ry)^\ell,
 \ee
 such that the perturbation fields may have regular solutions. Here $\ell$ is an non-negative integer, it  is just the quantum number associated with the harmonics in the ring coordinates. This can be seen clearly once  we take the large ring radius limit $R\to\infty$ and relate the black ring with the boosted black string.
 In practice we do not have to solve the four perturbation equations on $F$'s to find the explicit forms of the perturbation functions. Instead, with the help of the above boundary condition (\ref{blackringBC}), we are able to obtain the quasinormal mode condition for the frequency $\omega$.

 We have two kinds of the perturbations, the charge perturbation and the gravitational perturbation.
The charge perturbation is defined by $F_v(y)\neq F_q(y)$, which describes the fluctuation with a net charge. The gravitational perturbation is defined by $F_v(y)=F_q(y)$, which describes density fluctuation.

\paragraph{Charge perturbation} For the charge perturbation, from the perturbation equations we obtain
\be
\omega_c=\frac{\sqrt{R^2-1}}{R}\Big[\sqrt{1-2b_0}\,\hat{m}- i(\hat{m}^2+ \ell)\Big],
\ee
where $\hat{m}=m/R$. Hence the charge perturbation is stable.

\paragraph{Gravitational perturbation}  Due to the special form of the metric ansatz, the vector-type perturbation can be studied only for the charged slowly rotating black hole. For the charged black ring and the charged slowly boosted black sting, the momentum densities other than $p_z$ are needed to consider the vector-type perturbation. Here we only consider the most interesting scalar-type gravitational perturbation. The quasinormal mode condition for the scalar-type gravitational perturbation is given by
  \be
 \begin{split}
 &(1-2b_0)R^9\omega^3+i(1-2b_0)R^6\sqrt{R^2-1}\Bigl(3a_0m^2+3i\sqrt{1-2b_0}mR)+a_0(3\ell-2)R^2\Bigl)\omega^2\\
 &\,-(1-2b_0)R^3(R^2-1)\Bigl((3-6b_0+2b_0^2)m^4+6ia_0\sqrt{1-2b_0}m^3R+2m^2R^2((3-6b_0+2b_0^2)\ell\\
 &\,-2(2-4b_0+b_0^2))+2ia_0\sqrt{1-2b_0}(3\ell-2)mR^3+(3-6b_0+2b_0^2)\ell(\ell-1)R^4 \Bigl)\omega\\
 &\,+i(R^2-1)^{3/2}\Bigl(-a_0(1-2b_0)^2m^6+i\sqrt{1-2b_0}(-3+12b_0-14b_0^2+4b_0^3)m^5R\\
 &\quad +3(2b_0-1)(1-3b_0+2b_0^2)(\ell-2)m^4R^2-2i(1-2b_0)^{3/2}(3-6b_0+2b_0^2)(\ell-1)m^3R^3\\
  &\quad +i \sqrt{1-2b_0}(-3+12b_0-14b_0^2+4b_0^3)\ell(\ell-1)mR^5-a_0(1-2b_0)^2\ell^2(\ell-1)R^6\\
   &\quad-a_0(1-2b_0)^2(4-7\ell+3\ell^2)m^2R^4 \Bigl)=0,
 \end{split} \label{ringqnm}
 \ee
where we have introduced $a_0$ and $b_0$ to simplify the above expressions
 \be\label{a0b0}
 a_0=\frac{1+\sqrt{1-4Q^2}}{2},\quad b_0=\frac{1-\sqrt{1-4Q^2}}{2}.
 \ee
The neutral version  of the above equation, with  $a_0=1$ and $b_0=0$, is reduced to the quasinormal mode condition found in \cite{BRlargeD}. Note that the extremal case corresponds to $a_0=b_0=1/2$. The equation (\ref{ringqnm}) is a cubic equation and can be solved analytically. Qualitatively, the quasinormal modes are quite similar to the  ones of the neutral ring. For $\ell\neq0$ and $m\neq0$, all modes are  stable.

\begin{figure}[t]
 \begin{center}
  \includegraphics[width=65mm,angle=0]{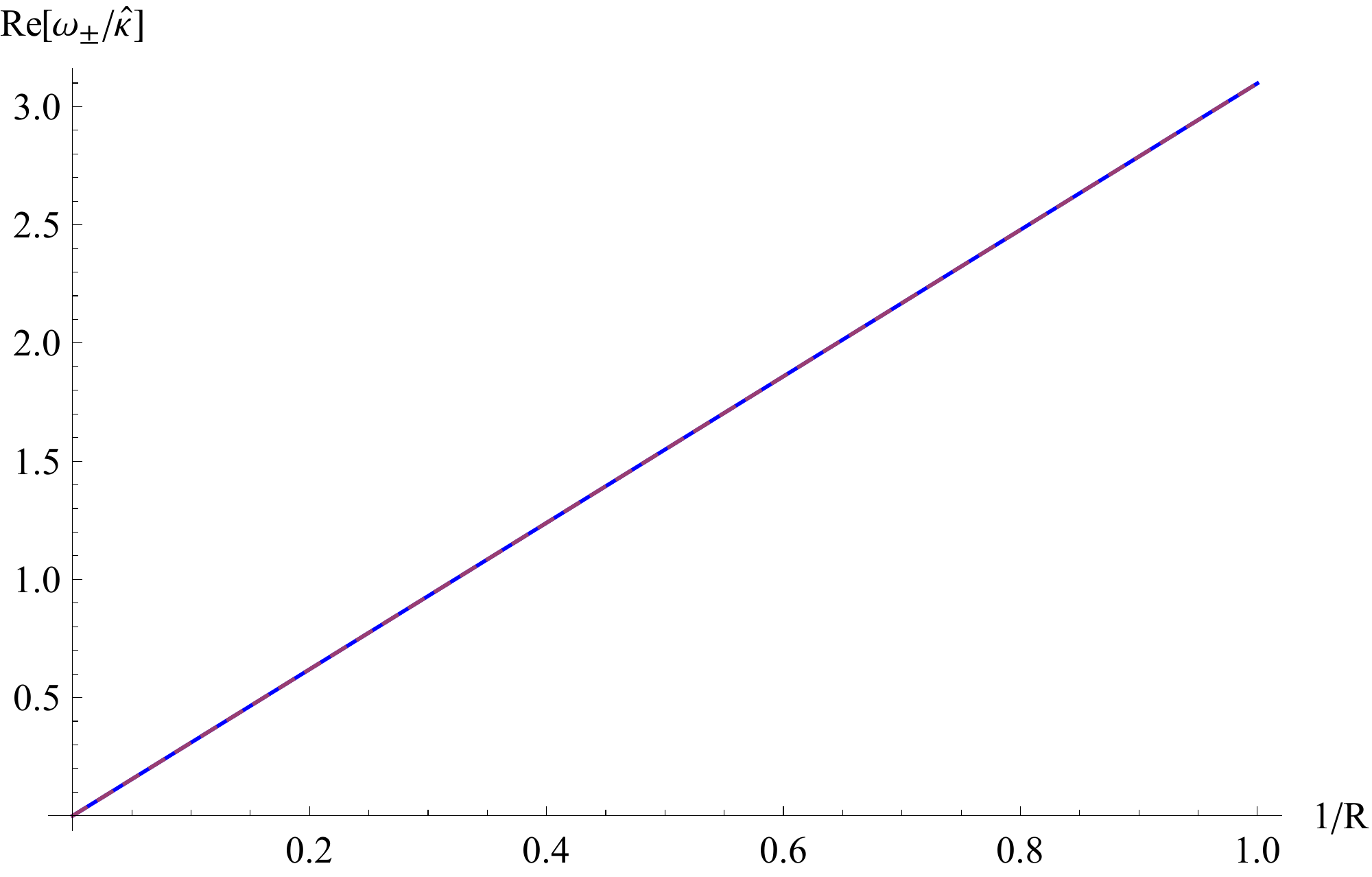}
  \hspace{5mm}
  \includegraphics[width=65mm,angle=0]{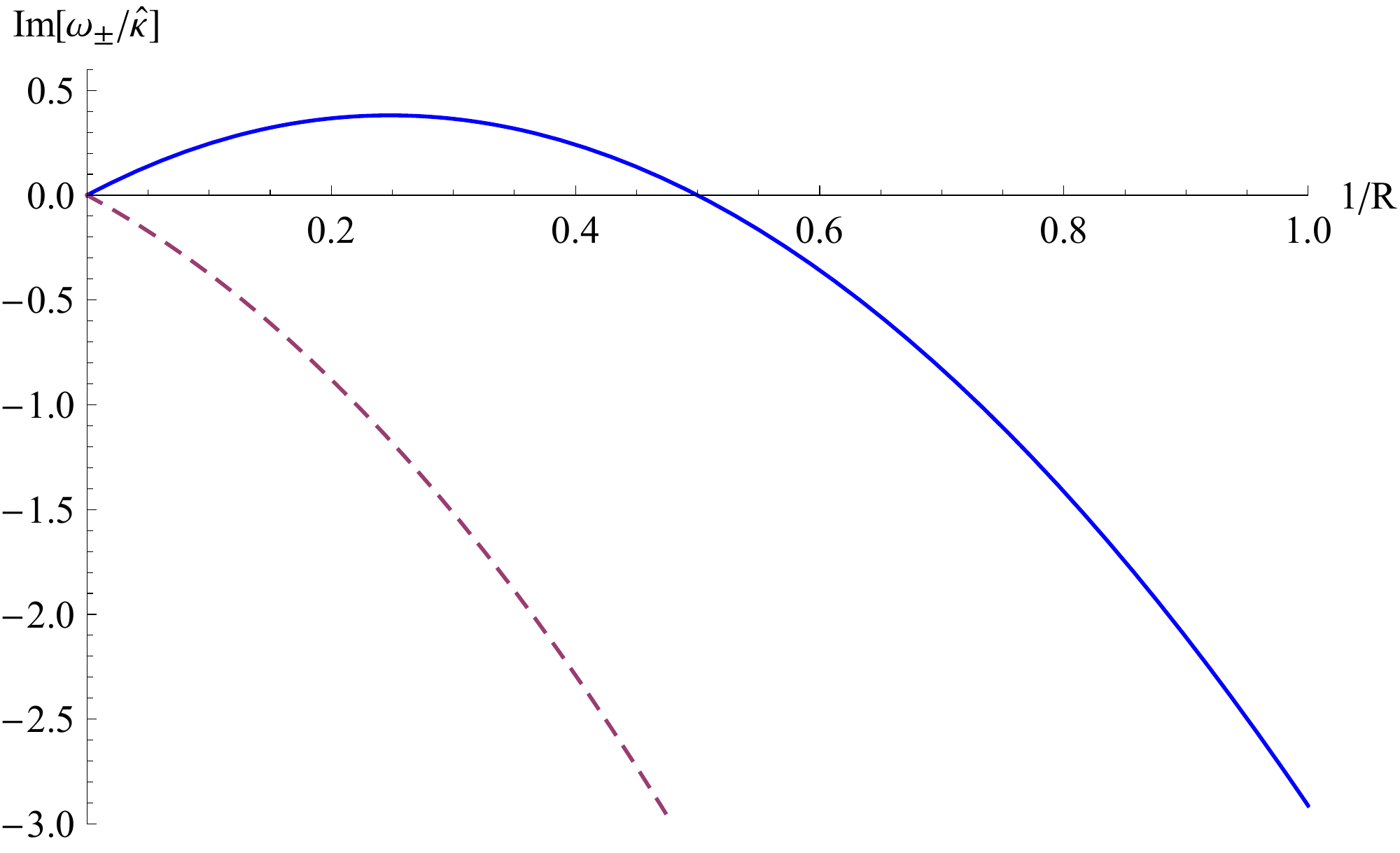}
 \end{center}
 \vspace{-5mm}
 \caption{The  frequencies $\omega_{+}^{(\ell=0)}$ (solid line) and $\omega_{-}^{(\ell=0)}$ (dashed line) of the scalar-type quasinormal modes of the charged black ring with $\ell=0$, $m=2$ and $b_0=0.2$ . The real part and the imaginary part are depicted in the left and right panel respectively, both of which are in unit of the reduced surface gravity $\hat{\kappa}$. $\mathrm{Im}[\omega_{+}^{(\ell=0)}]$ is positive for $R>2$ which signals an GL-like instability.}
 \label{fig1}
\end{figure}

  For $\ell=0$ and $m\neq0$, the quasinormal condition can be solved by
 \be
 \omega_0^{(\ell=0)}=\frac{\sqrt{R^2-1}}{R}\Bigl[\hat{m}-ia_0(\hat{m}^2-2)\Bigl],
 \ee
 and
 \be\label{BlackringQNMl=0}
 \omega_{\pm}^{(\ell=0)}=\frac{\sqrt{R^2-1}}{R}\Bigl[\sqrt{1-2b_0}\hat{m}\pm i\hat{m}(\sqrt{a_0-b_0+b_0^2\, \hat{m}^2}\mp a_0\hat{m})\Bigl],
 \ee
 The mode with $\omega_0^{(\ell=0)}$ can be regarded as a gauge mode, since when taking the large radius limit $R\gg1$, the corresponding counterpart of the boosted black string does not exist. On the contrary as we will see later, the modes with $\omega_{\pm}^{\ell=0}$ have correspondents in the boosted black string. The real and imaginary parts of the quasinormal modes $\omega_{\pm}^{(\ell=0)}$ are plotted in Figure \ref{fig1}, with $m=0$ and $b_0=0.2$.

Identical to the neutral case \cite{BRlargeD}, the real part of $\omega_{+}^{(\ell=0)}$ always saturates the superradiant condition
 \be
 \text{Re}\Big[ \omega_{+}^{(\ell=0)}\Big]=m\hat{\Omega}_H,
 \ee
where $\hat{\Omega}_H$ is defined by (\ref{balancecondition}). The same relation  has also been found in the equally rotating and singly rotating  Myer-Perry black holes at large $D$ \cite{EST:rotating, StationarylargeD, Tanabe:chargedrotating}. However, by adding the $1/n$ correction, it turns out that above condition  does not hold any more. We expect the same thing happens for the charged case.

When $R>m$,  $\omega_{+}^{(\ell=0)}$ has a positive imaginary part, indicating the  instability. It is interesting to observe that the critical ring radius  $R=m$ at which  $\mathrm{Im}[\omega_{+}^{(\ell=0)}]$ vanishes is independent of the charge.

\begin{figure}[t]
 \begin{center}
  \includegraphics[width=65mm,angle=0]{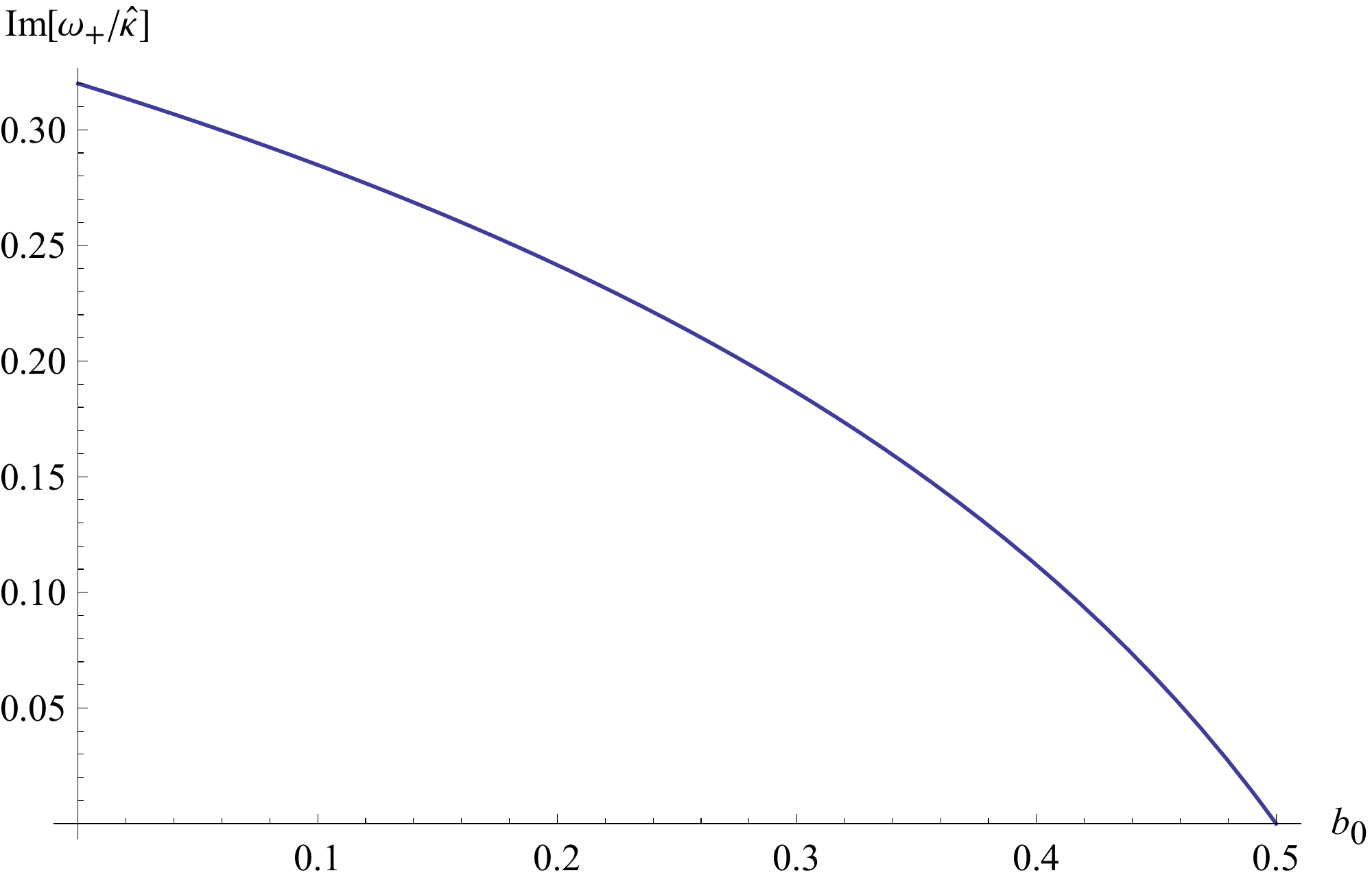}
 \end{center}
 \vspace{-5mm}
 \caption{Dependence of $\mathrm{Im}[\omega_{+}^{(\ell=0)}]$  of the charged black ring on the electric charge with $\ell=0$, $m=2$ and $R=1/2$.}
 \label{fig2}
\end{figure}

The presence of electric charge makes the GL-like instability weaker. From Figure \ref{fig2}, it is easy to find that a larger $b_0$ leads to a smaller $\mathrm{Im}[\omega_{+}^{(\ell=0)}]$.  This effect is different from the one in the charged MP black hole \cite{Tanabe:chargedrotating} constructed by taking the large $D$ boost transformation between equally rotating MP black hole and Schwarzschild black hole. Here the charge and the angular velocity $\hat{\Omega}_H$  is related by (\ref{balancecondition}), a larger charge means a smaller angular momentum via (\ref{angularmomentum}), but in \cite{Tanabe:chargedrotating} the rotation and the charge are independent of each other, so the electric charge can enhance the ultraspinning instability.

At the extremal case, $b_0=a_0=1/2$ such that $\omega_{+}^{(\ell=0)}=0$ and the black ring seems to become marginally unstable.  The similar  phenomenon  appears in the discussion of boosted black string as well, see section \ref{chargedslowlystring}.

For the axisymmetric  modes $m=0$,  the QNM condition is solved by
\be
\omega_0^{(m=0)}=-i\frac{\sqrt{R^2-1}}{R}a_0\ell,
\ee
and
\be
\omega_{\pm}^{m=0}=\frac{\sqrt{R^2-1}}{R}\Bigl[-i a_0\, (\ell-1)\pm\sqrt{(\ell-1)(a_0^2-b_0^2 \ell)}\Bigl].
\ee
The frequencies of the quasinormal modes  are the same as the ones of the large $D$ RN black hole\footnote{The charge introduced in \cite{chargedmembrane, TanabeIndS} is slightly different from ours.  In their notations, $a_0=1/(1+Q^2)$ and $b_0=Q^2/(1+Q^2)$. }\cite{chargedmembrane, TanabeIndS} up to a  multiplicative factor characterizing the finiteness of the ring. This is  easy to understand since for $m=0$ the perturbation has no dependence on $\phi$, so the situation is just like a $(D-1)$-dimensional RN black hole, with the only difference being that the $\phi$ direction is compact now.

Note that the mode $\omega_0^{(m=0)}$ should correspond to the vector gravitational perturbations of the RN black hole. The reason for this  is similar to the one for the MP black hole \cite{EST:rotating}. When discussing the perturbation of the RN black hole we usually use vector harmonics on $S^{D-2}$ with  the  angular momentum number $\ell_v$. The subtlety is that the vector harmonics on $S^{D-2}$  can be decomposed into two scalar harmonics on $S^{D-4}$, with the corresponding angular momentum number $\ell=\ell_v\pm1$. So the $\omega_0$ mode is just the vector-type gravitational perturbations of the RN black hole with $\ell=\ell_v-1$.

Like the neutral case, the axisymmetric perturbation does not show any instability at leading order of $1/n$. However, numerical study \cite{SW, FKT} showed that   for neutral fat black ring in five dimensions the axisymmetric perturbations are unstable. It is natural to expect such phenomena exists for the charged black ring as well.  The reason why the large $D$ expansion can not capture this feature is not clear at present.

\subsection{Black string limit}\label{blackstringlimit}
It is well known that  when the ring radius $R\gg1$, that is the  black ring is very thin, the black ring is identical to a boosted black string \cite{EHNOR}. In the following we
show this point clearly. At the large ring radius limit, from (\ref{Py}) $P(y)$ behaves like
\be
P(y)=\mc O(1/R),
\ee
by setting $P_0=P_1=0$. The black ring solution (\ref{ringsolution}) at the large radius limit becomes
\be
 \begin{split}
ds^2&=-\Big(1-\frac{1}{\sR}+\frac{Q^2}{\sR^2}\Big)dv^2+2dvdr-2\sqrt{\frac{1-2b_0}{n}}\Big(\frac{1}{\sR}-\frac{Q^2}{\sR^2}\Big)dvdx\\
&\,+\Big[1+\frac{(1-2b_0)}{n}\Big(\frac{1}{\sR}-\frac{Q^2}{\sR^2}\Big)-\frac{2}{n}\ln \Big(1-\frac{b_0}{\sR}\Big)\Big]dx^2+r^2dz^2+r^2d\Omega_n^2,
 \end{split}\label{largeradius}
\ee
and the gauge field (\ref{ringsolutiongaugefield}) becomes
\be
A_\mu dx^\mu=\frac{\sqrt{2}Q}{\sR}dv
-\sqrt{\frac{1-2b_0}{n}}\frac{\sqrt{2}Q}{\sR}dx,
\ee
where the string direction $dx$  is identified by $dx=Rd\Phi=Rd\phi/\sqrt{n}$. Under such large radius limit, the above expressions are just the ones of the  boosted black string at large $D$
 (see (\ref{boostedblackstring}))
with the boosted velocity $\alpha$ given by
\be\label{boostvelocity}
\mathrm{sinh}\,\alpha=\sqrt{\frac{1-2b_0}{n}}.
\ee
It is obvious when there is charge, the boost transformation  depends on it and at the extremal limit the boost velocity becomes zero. We can use the boost relation  to associate the large radius limit of the QNMs $\omega_{\pm}^{(\ell=0)}$ of the charged black ring with the ones of the charged black string. The QNMs of the charged black string were obtained in \cite{EILT:Hydro} by using the large $D$ effective theory. In section \ref{chargedslowlystring} we give the direct derivation of the QNMs of the boosted string from
the effective equations, and show that they are related to the ones of \cite{EILT:Hydro} by a simple boost transformation. For the S-wave sector of the gravitational perturbation $\sim e^{-i\omega v+ikx}$, the QNMs are given by
\be
\omega_{\pm}^{st}=\pm i \hat{k}\sqrt{\hat{k}^2 b_0^2+a_0-b_0}-i a_0 \hat{k}^2,
\ee
where $\hat{k}=k/\sqrt{n}$ is the quantum number  with respect to $\phi$.  The corresponding QNMs of the boosted black string is
obtained by performing such boost transformation
\be
dv\to \mathrm{cosh}\,\alpha\, dv-  \mathrm{sinh}\,\alpha\,dx,\qquad dx\to \mathrm{cosh}\,\alpha\, dx-  \mathrm{sinh}\,\alpha\,dv,
\ee
with the boost velocity given by (\ref{boostvelocity}). This leads to
\be
\omega_{\pm}^{bst}=k\sinh\,\alpha+\omega_{\pm}^{st} \cosh\,\alpha.
\ee
Taking into account of the identification $\hat{k}=\hat{m}$, the above frequencies are in accord with the ones of the QNMs of the charged black ring (\ref{BlackringQNMl=0}) in the large radius limit.

\section{Other Charged Slowly Rotating Black Objects}\label{chargedalowlyBH}

The effective equations in the large $D$ expansion found in section \ref{EffEqs} can actually describe other charged slowly rotating black objects,
including the black holes and the black strings. In this section, we discuss these two cases briefly. The essential point in the construction is to determine the  functions $G(z)$ and $H(z)$ in different cases. %In particular, the charged rotating black holes have only  been investigated in the slowly rotating limit\cite{Aliev}, in which the rotation is taken as a perturbation. The assumption (\ref{angualrvecofthinring}) fits well with this spirit.

\subsection{Charged Slowly Rotating Black Holes}\label{chargedalowlyMP}

To discuss the charged slowly rotating black hole, we need to consider the metric of $D=n+4$ dimensional flat spacetime  in spherical coordinate
\be
ds^2=-dt^2+dr^2+r^2(dz^2+\mathrm{sin}^2\,z\,d\Phi^2+\mathrm{cos}^2\,z\,d\Omega_n^2).
\ee
As before, the embedding of the leading order metric is obtained by taking $r=1$. This leads to the identification
\be\label{slowlyrotEM}
H(z)=\mathrm{cos}\,z,\quad G(z)=\mathrm{sin}\,z.
\ee
It is clear that the relation (\ref{momentumcons}) is satisfied. From (\ref{hatkappa}) and (\ref{surfacegra}) we get
\be
\hat{\kappa}=\frac{1}{2},\quad \kappa=\frac{n}{2}\frac{\sqrt{1-4Q^2}+4Q^2-1}{2Q^2}.
\ee
\paragraph{Stationary solution} Directly from (\ref{EffEq2r}), (\ref{Stationarysolutionq}) and (\ref{Stationarysolutionpphi}) we obtain
 \be\label{slowrotstat}
 q(z)=Q\, p_v(z),\quad p_z(z)=p_v'(z),\quad p_\phi(z)= \hat{a}\, \mathrm{sin}^2\,z\, p_v(z),
 \ee
 where in order to compare with the familiar rotation parameter we introduce $\hat{a}=\hat{\Omega}_H$ in (\ref{Stationarysolutionpphi}). Furthermore $p_v(z)=e^{P(z)}$ can be determined by (\ref{EffEq3r}). For $1-4Q^2\neq0$ after  substituting (\ref{slowrotstat}) and (\ref{slowlyrotEM}) into (\ref{EffEq3r}) we have
  \be
  P''(z)+P'(z)\, \mathrm{tan}\,z-\frac{\hat{a}^2\mathrm{cos}^2\,z}{\sqrt{1-4Q^2}}P(z)=0,
  \ee
  whose solution  is
  \be\label{Pz}
  P(z)=p_0+d_0\, \mathrm{sin}\,z-\frac{\hat{a}^2\mathrm{cos}^2\,z}{2\sqrt{1-4Q^2}}.
  \ee
Similar to the uncharged case in \cite{BRlargeD}, here the integration constants $p_0$ and $d_0$ describe trivial deformations of the solution, so we set them to be zero. Then we completely determine the stationary solution of the effective equations.

The metric and the gauge field of the solution are of the following form
  \be
   \begin{split}
  ds^2=&-\Big(1-\frac{p_v(z)}{\sR}+\frac{Q^2p_v(z)^2}{\sR^2}\Big)dv^2+2dvdr-2a\sin^2z\,\Big(\frac{p_v(z)}{\sR}-\frac{Q^2p_v(z)^2}{\sR^2}\Big)dvd\Phi\\
  &-\frac{2\,p_v'(z)}{n}\Big(\frac{1}{\sR}-\frac{Q^2p_v(z)}{\sR^2}\Big)dvdz+r^2\sin^2z\Big[1+a^2\sin^2z\,\Big(\frac{p_v(z)}{\sR}-\frac{Q^2p_v(z)^2}{\sR^2}\Big)\Big]d\Phi^2\\
  &+\frac{2r^2}{n}a\,p_v'(z)\sin^2z\Big(\frac{p_v(z)}{\sR}-\frac{Q^2p_v(z)^2}{\sR^2}\Big)dzd\Phi+r^2dz^2+ r^2\cos^2z\, d\Omega_n^2,
   \end{split}
  \ee
  and
  \be
  A_\mu dx^\mu=\frac{\sqrt{2}Qp_v(z)}{\sR}dv-\frac{1}{n}\frac{\sqrt{2}Qp_v'(z)}{\sR}-\frac{\sqrt{2}a\sin^2z\,p_v(z)}{\sR}d\Phi.
  \ee
  This solution describes the $D=n+4$ dimensional charged slowly rotating black hole in the Einstein-Maxwell theory, with $a=\hat{a}/\sqrt{n}$. When $\hat{a}=0$, it is easy to see that the solution is reduced to the RN black hole, which is expected. When $Q=0$, that is the neutral case, the solution can be related to the slowly rotating MP black hole by a coordinate transformation.  Due to the fact that
 there is no  exact form of  the charged rotating black hole, namely the analog of Kerr-Newman solution,  in the dimensions higher than four, it seems impossible to check the validness of
our solution. However, in \cite{Aliev}, the charged rotating black hole with a single small angular momentum has been discussed. In the obtained solution,  only the linear order term in  $a$ appears in the metric and the higher order terms are omitted. At the leading order in the $1/n$ expansion and keep leading order of $a$, the solution we constructed above simply reproduces the solution in \cite{Aliev}. This strongly supports that we have obtained the correct charged slowly rotating black hole solution in the Einstein-Maxwell theory with $a\sim \mc{O}(1/\sqrt{n})$.

%The discussion on the effective equations at large $D$  for the charged rotating black holes becomes subtle in the extremal limit $1-4Q^2=0$, from (\ref{Effextremal}), it gives requirement
The discussion on the effective equation at large $D$ for the charged rotating black holes
becomes subtle in the extremal limit. In this limit, $1-4Q^2\to 0$, $P(z)\to -\infty$,  we would have $p_v=q=p_z=p_\phi=0$, which leads to a  flat metric and a vanishing gauge field. This is not a reasonable result. The reason why this phenomenon occurs is that when doing  the  $1/n$ expansion we need to specify the large $D$ scaling of every function in the metric and the gauge potential. As discussed in section \ref{setup}, we have already assumed that $P(z)=\mc O(1)$ (note that both $\hat{a}=0$ and  $Q=0$ belong to this situation),  then for consistence we should require that $1-4Q^2$ is of $\mc O(1)$. When approaching the extremal limit like $\sqrt{1-4Q^2}=e/n$, where $e$ is an quantity of $\mc O(1)$, then  $p_v(z)$ becomes like $e^{-n k(z) }$, where $k(z)$ is a function of $z$ and is of  $\mc O(1)$, and the  $1/n$ expansion breaks down. We may  make an estimation for the validness of the  $1/n$ expansion.  We should demand that
 \be
 \frac{\hat{a}^2}{\sqrt{1-4Q^2}}\ll n,
 \ee
in order to keep $P(z)=\mc O(1)$. In other words, the charge $Q$ is required to satisfy
\be
1-4Q^2\gg \frac{\hat{a}^4}{n^2}.\label{extremecon}
\ee
This condition gives a more restrictive constraint on the charge than (\ref{chargecondition}). When $Q=0$, the above condition becomes
\be
\hat{a}\ll\sqrt{n},
\ee
this is consistent with the previous  assumption that the horizon angular  velocity is of $\mc O(1/\sqrt{n})$.  Physically, we can understand this phenomenon as follows. Even though the black hole is very slowly rotating, it provides a centrifugal force (emerges only at the second order of $a$), which could be tiny but could not
be neglected in the extremal limit. If we naively take the extremal limit $1-4Q^2=0$, the existence of the extra angular momentum makes the
black hole unphysical, breaking the cosmic censorship. In order to have a physically acceptable solution, we must have the condition (\ref{extremecon}).
Therefore, we may just assume that
\be\label{hata}
\hat{a}=(1-4Q^2)^{1/4}\,\hat{b},
\ee
where $\hat{b}$ is of order 1.  In terms of $\hat{b}$, $P(z)$ now becomes like
\be
P(z)=-\frac{\hat{b}^2\,\mathrm{cos}^2\,z}{2},
\ee
which is regular even in the extremal limit $\hat{b}=0$. The relation in (\ref{hata}) is similar to (\ref{balancecondition})  for the charged black ring.  The difference is that the solution here is regular at the event horizon, while the obtained charged black ring is
singular at the event horizon in the extremal limit.

\paragraph{Quasinormal modes}Now let us investigate the quasinormal modes of the charged slowly rotating black hole. First consider the following perturbation  around the aforementioned stationary solution
\be
\begin{split}
p_v(v,z,\phi)&=e^{-\frac{\hat{b}^2\,\mathrm{cos}^2\,z}{2}}\big(1+\epsilon F_v(z)e^{-i\omega v}e^{im\phi} \big),\\
q(v,z,\phi)&=Qe^{-\frac{\hat{b}^2\,\mathrm{cos}^2\,z}{2}}\big(1+\epsilon F_q(z)e^{-i\omega v}e^{im\phi} \big),\\
p_z(v,z,\phi)&=\hat{b}^2\mathrm{sin}\,z\,\mathrm{cos}\,z\,e^{-\frac{\hat{b}^2\,\mathrm{cos}^2\,z}{2}}\big(1+\epsilon F_z(z)e^{-i\omega v}e^{im\phi} \big),\\
p_\phi(v,z,\phi)&=(1-4Q^2)^{1/4}\,\hat{b}\,\mathrm{sin}^2\,z\,e^{-\frac{\hat{b}^2\,\mathrm{cos}^2\,z}{2}}\big(1+\epsilon F_\phi(z)e^{-i\omega v}e^{im\phi} \big).
\end{split}
\ee
Then plugging these ansatz into the effective equations, and taking in account that at $z=0$, $F_v(z)\propto z^\ell$, which stems from the large $D$ behavior of spherical harmonics on $S^{n+1}$ \cite{EST:rotating}, we can determine the frequencies of the quasinomal modes. Moreover, for simplicity we focus on the case  $m=\mc O(1/n)$,  A more  convenient quantity $\bar{m}$ is defined by
 \be
\bar{m}=nm,
\ee
which now is of order $\mc O(1)$. Then the QNM frequencies are characterized by $\ell$, $\bar{m}$ and $\hat{b}$.

\paragraph{Charge perturbation} For the charge perturbation, from the perturbation equations we obtain
\be
\omega_c=- i \ell,
\ee
which suggest that the perturbation is stable.
\paragraph{Vector-type gravitational perturbation} For the charged rotating black hole, it is possible to discuss the vector-type gravitational perturbation. We find the QNM frequency
\be
\omega_v=-ia_0\ell.
\ee
\paragraph{Scalar-type gravitational perturbation} For the scalar-type gravitational perturbation, we have
\be
\omega_0=-i a_0 (\ell-2), \quad  \omega_{\pm}=-i a_0\, (\ell-1)\pm\sqrt{(\ell-1)(a_0^2-b_0^2 \ell)},
  \ee
  with $a_0$ and $b_0$ being defined in (\ref{a0b0}). Note that the modes with the frequency $\omega_v$ and $\omega_0$ correspond to the vector-type gravitational perturbations of the RN black hole with $\ell=\ell_v-1$ and $\ell_v+1$ respectively, where $\ell_v$ is the  the  angular momentum number of the vector harmonics on $S^{D-2}$.
The modes with $\omega_{\pm}$ reproduce the QNMs of the RN black hole \cite{TanabeIndS, chargedmembrane}. This  is expected since we are considering a slowly rotating black hole with $m=\mc O(1/n)$, at leading order  the dependence of the perturbation on $\phi$ disappears and the QNMs are captured by the ones in the  RN black hole with one lower dimensions. At the next order in $1/n$, like the neutral case \cite{BRlargeD} $\bar{m}$ will appear in the QNMs.

As observed in \cite{Tanabe:chargedrotating}, the charge can enhance the ultraspinning instability of rotating MP black holes.  Even the rotation is small but the charge is large, the black hole can still be unstable. However we cannot find this behavior in the solution we found.  The reason is that here we consider $a=\mc O(1/\sqrt{n})$ hence the effect of the rotation is pushed to  the next order, as demonstrated in the neutral case \cite{BRlargeD}. In order to see the effect of the charge on the instability we should consider the case with $a=\mc O(1)$, the discussion of which should be more complicated.   Nevertheless, we may still see some indications. When the charge is large enough, the frequency $\omega_+$ becomes imaginary and increases with the charge.  When $a_0-b_0=0$, $\omega_+ = 0$, so this solution branch has potential to develop instability if adding the rotation. The similar phenomenon has been found for equally rotating MP black holes \cite{Tanabe:chargedrotating}. In this sense,  the charge may help to destabilize the rotating black hole.

\subsection{Charged slowly boosted black string}\label{chargedslowlystring}

To discuss the black string, we consider the embedding in the spherical coordinate.  The metric of $D=n+4$ dimensional spacetime  with one compact direction in the spherical coordinate is
\be
ds^2=-dt^2+d\Phi^2+dr^2+r^2(dz^2+\mathrm{sin}^2 z\, d\Omega^2_n),
\ee
where $\Phi$ is the coordinate of the compact direction. The embedding of $r=1$ of the leading order metric into the above background gives the  following identifications
\be
G(z)=1,\quad H(z)=\mathrm{sin}\,z.
\ee
Since the embedded solution has $S^1\times S^{n+1}$ horizon topology, the effective equations describe the non-linear dynamical deformations of the charged slowly boosted black string.
For stationary solution we have
\be
p_v(z)=e^{P(z)},\quad p_\phi=p_\phi(z),\quad p_z=p_z(z).
\ee
\paragraph{Stationary solution} Directly from (\ref{EffEq2r}), (\ref{Stationarysolutionq}) and (\ref{Stationarysolutionpphi})  we have
 \be
q(z)=Q p_v(z),\quad  p_\phi(z)=\hat{\sigma} p_v(z),\quad p_z(z)=p_v'(z).
 \ee
Here $\hat{\sigma} $ is an integration constant  describing the boost effect of the black string.  For $1-4Q^2\neq0$, the equation of $P(z)$ is given by
 \be
 P''(z)-\mathrm{cot}\,z\, P'(z)=0,
 \ee
which can be solved easily and leads to
\be
p_v(z)=e^{p_0+p_1 \mathrm{cos}\,z}.
\ee
 The integration constants $p_0$ and $p_1$ are trivial deformations:
$p_0$ is the $1/n$ redefinition of $r_0$, and $p_1$ is the redefinition of $z$ coordinate. We may  set them to be zero.
Then we can write down the leading order metric
\be
\begin{split}
ds^2=&-\Big(1-\frac{1}{\sR}+\frac{Q^2}{\sR^2}\Big)dv^2+2dvdr+\frac{\hat{\sigma}}{\sqrt{n}} \Big(\frac{1}{\sR}-\frac{Q^2}{\sR^2}\Big)dvd\Phi\\
&+\biggl[1+\frac{\hat{\sigma}^2}{n}\Big(\frac{1}{\sR}-\frac{Q^2}{\sR^2}\Big)-\frac{2}{n}\ln\Big(1-\frac{b_0}{\sR}\Big)\biggl]d\Phi^2+r^2 dz^2+r^2 d\Omega_n^2,
\end{split}\label{boostedblackstring}
\ee
and the gauge field
\be
A_\mu dx^\mu=\frac{\sqrt{2}Q}{\sR}dv
-\frac{\hat{\sigma}}{\sqrt{n}}\frac{\sqrt{2}Q}{\sR}d\Phi.
\ee
The metric is related to the one in \cite{CEV:Higherrot, EILT:Hydro, RE:OnBrane} by a boost transformation with the boost velocity $\b$
\be
\mathrm{sinh}\,\b=\frac{\hat{\sigma}}{\sqrt{n}}.
\ee
Comparing (\ref{boostedblackstring}) with (\ref{largeradius}), we see that when the boost velocity of the charged boosted black string takes a specific value, that is $\beta=\alpha$, where $\alpha$ is defined by (\ref{boostvelocity}), then the large radius limit of the charged black ring is reproduced.

For $1-4Q^2=0$, (\ref{Effextremal}) is satisfied automatically since $G'(z)=0$ so we do not find any constraint on the solution in the
extremal limit. As we mentioned before, the solution (\ref{boostedblackstring}) has a singularity at the event horizon, from \cite{CEV:Higherrot}
we know that this singularity stems from the behavior of the solution itself.

\paragraph{Non-uniform charged black string}Another stationary solution of the effective equations is the non-uniform black string which is believed to be the endpoint of the evolution of the large $D$ charged black string, as showed in \cite{EILT:Hydro,ST:Nonuni}. The non-uniform means the solution is inhomogeneous along the $\Phi$ direction. In this case, there is only one Killing vector $\partial_v$. Moreover, we do not need the $z$ dependence anymore, so $p_z$ can be set to zero. We have
\be
p_v=p_v(\phi), \quad q=q(\phi),\quad p_\phi=p_\phi(\phi).
\ee
Then from the effective equations we have
\be
q=Q p_v(\phi), \quad  p_\phi=p_v'(\phi),
\ee
and
\be
\partial_\phi^3 p_v+\partial_\phi p_v-\frac{2\partial_\phi p_v \partial_\phi^2 p_v}{p_v}+\frac{(\partial_\phi p_v)^3}{p_v^2}=0.
\ee
This equation can be solve numerically, but we will not pursue this anymore.

\paragraph{Quasinormal modes}Let us now investigate the quasinormal modes of the charged boosted black string we constructed above. Considering the perturbations around the stationary solution, 
\be
\begin{split}
p_v(v,z,\phi)&=1+\epsilon F_v(z)e^{-i\omega v+i\hat{k}\phi},\\
q(v,z,\phi)&=Q\big(1+\epsilon F_q(z)e^{-i\omega v+i\hat{k}\phi} \big),\\
p_z(v,z,\phi)&=\epsilon F_z(z)e^{-i\omega v+i\hat{k}\phi},\\
p_\phi(v,z,\phi)&=\hat{\sigma}\big(1+\epsilon F_\phi(z)e^{-i\omega v+i\hat{k}\phi} \big).
\end{split}
\ee
As before after plugging these expressions into the effective equations and taking into account that at $z=0$, $F_v(z)\propto z^\ell$, then
we obtain the QNMs for the charge perturbation and the scalar type gravitation perturbation respectively as follows.
\paragraph{Charge perturbation} The QNM frequency for this type perturbation is
\be
\omega_c=\hat{k}\hat{\sigma}-i(\hat{k}^2+\ell).
\ee
Obviously, the charge perturbation is stable.
\paragraph{Gravitational perturbation} For the scalar-type gravitational perturbation, we
obtain the quasinormal mode condition
\be
\begin{split}
\,&\omega^3+\omega^2\Big[(-2i+3i \hat{k}^2+3i\ell)a_0-3\hat{\sigma}\hat{k}\Big]+\omega\Big[\ell^2(1-2a_0^2-2a_0)+\ell\big(-1+2\hat{k}^2+a_0^2(2-4\hat{k}^2)\\
&+a_0(2-4\hat{k}^2-6ik\hat{\sigma})\big)+\hat{k}^4(1-2a_0^2-2a_0)-6ia_0 \hat{k}^3\hat{\sigma}+\hat{k}^2(-1+2a_0+4a_0^2+3\hat{\sigma}^2)\\
&+4ia_0\hat{k}\hat{\sigma}\Big]+i\ell^3a_0(1-2a_0)+\ell^2\Big(-\hat{k}\hat{\sigma}+2a_0^2(i-3i\hat{k}^2+\hat{k}\hat{\sigma})+a_0(-i+3i\hat{k}^2+2\hat{k}\hat{\sigma})\Big)\\
&+\ell \hat{k}\big(\hat{\sigma}-2\hat{k}^2\hat{\sigma}+a_0^2(8i\hat{k}-6i\hat{k}^3-2\hat{\sigma}+4\hat{k}^2\hat{\sigma})+a_0\big(3i\hat{k}^3-2\hat{\sigma}+4\hat{k}^2\hat{\sigma}
+i\hat{k}(-4+3\hat{\sigma}^2)\big)\Big)\\
&+\hat{k}^2\Big(2a_0^2(\hat{k}^2-2)(i-i\hat{k}^2+\hat{k}\hat{\sigma})+ia_0\big(2+\hat{k}^4+2i\hat{k}\hat{\sigma}-2i\hat{k}^3\hat{\sigma}-2\hat{\sigma}^2+3\hat{k}^2(-1+\hat{\sigma}^2)\big)\\
&-\hat{k}\hat{\sigma}(-1+\hat{k}^2+\hat{\sigma}^2)\Big)=0.
\end{split}
\ee
When $a_0=1$,  the above equation is reduced to the quasinormal mode condition of the slowly boosted black string \cite{BRlargeD}. For $\ell=0$, this cubic equation of $\omega$ is easy to solve,
\be\label{boostedstrinQNMl=0}
\omega_0^{(\ell=0)}= \hat{k}\hat{\sigma}-ia_0(\hat{k}^2-2), \quad \omega_{\pm}^{(\ell=0)}=\hat{k}\hat{\sigma}\pm i\hat{k}\big(\sqrt{b_0^2\,\hat{k}^2+a_0-b_0}\mp a_0\hat{k}\big).
\ee
The mode $\omega_0^{\ell=0}$ should be discarded as it is a gauge mode. This is reflected in the fact  that when $a_0=1$ and $\hat{\sigma}=0$,  the direct calculation from the metric perturbation in \cite{EST:lageDlimit, EST:QNMAdS} shows that no such counterpart exists. The remaining two solutions are physical and correspond to the quasinormal modes of the S-wave sector of the scalar type gravitational perturbation of the charged  boosted black string. When $\hat{\sigma}=0$, the above result reproduces the one of the charged black string obtained in \cite{EILT:Hydro}. The $\omega_{+}^{\ell=0}$ solution branch has a positive imaginary part for
\be
\hat{k}<1,
\ee
which signals the Gregory-Laflamme instability \cite{GL, GLcharge}. It turns out that the charge has no effect on the critical wavenumber ($\hat{k}_c=1$) of  the threshold mode.
Moreover, the imaginary part of $\omega_{+}^{(\ell=0)}$ decreases for a larger $b_0$,  indicating that the presence of charge weakens the instability.

For $\hat{k}=0$, the quasinormal normal condition leads to
\be
\omega_0^{(\hat{k}=0)}=-i a_0\ell,\quad w_{\pm}^{(\hat{k}=0)}=-i a_0(\ell-1)\pm\sqrt{(\ell-1)(a_0^2-b_0^2\ell)}.
\ee
Similar to the analysis for  axisymmetric perturbation ($m=0$) of the charged black ring,  these modes correspond to the decoupled vector and scalar modes of RN black hole on $S^{D-2}$.
For $\ell\neq 0$ and $\hat{k}\neq0$, no unstable mode is found. Therefore for the charged boosted black string, similar to the neutral case\cite{Kudoh:Origin}, the instability only exists in the S-wave ($\ell=0$) sector.

As we explained above, the charged boosted black string is related to the large radius limit of charged black ring once the boost velocity is specified to be  $\hat{\sigma}=\sqrt{1-2b_0}$. The quasinormal modes on both sides should also be connected by such relation. Indeed, if we take the large radius limit of the quasinormal modes $\omega_{\pm}^{\ell=0}$ of the charged black ring in (\ref{BlackringQNMl=0}), we obtain
\be
 \omega_{\pm}^{(\ell=0)}=\sqrt{1-2b_0}\hat{m}\pm i\hat{m}\Big(\sqrt{a_0-b_0+b_0^2\, \hat{m}^2}\mp a_0\hat{m}\Big).
 \ee
By identifying $\hat{k}=\hat{m}$, and using $\hat{\sigma}=\sqrt{1-2b_0}$, we find that the result is identical to (\ref{boostedstrinQNMl=0}).
This relation holds for other modes as well.

\section{Summary}\label{summary}

In this paper, by using the large $D$ effective theory of the black hole, we studied the charged   black holes in the Einstein-Maxwell theory. These black holes include the charged  black ring, the charged slowly rotating MP black hole and the charged slowly boosted black string. According to the property of thin black ring obtained by the blackfold method at large $D$ and the effect of the charge on the string tension, it is believed that the large $D$ charged black ring belongs to the class of slowly rotating black holes with the  horizon angular velocity being of $\mc O (1/\sqrt{D})$. Consequently the rotation can be treated  perturbatively, which makes the construction of the charged black ring possible.

After integrating out the radial direction of the Einstein-Maxwell equations, we obtained the effective equations for the charged slowly rotating black holes. The charged black ring were constructed analytically as the stationary solution with the embedding  into flat spacetime background  in the ring coordinate. We found  that the presence of the charge lowers the horizon angular velocity and the angular momentum, which is consistent with the intuition that the charge provides repulsive force to against the string tension. By performing perturbation analysis of the effective equations  we obtained the QNMs of the charged black ring. We found the GL-like instability  for the non-axisymmetric perturbation when the charged black ring is relatively thin. Moreover we noticed that the charge weakens the GL-like instability.  As byproducts, we  constructed the charged slowly rotating MP black hole and the charged slowly boosted black string via different embeddings, and studied their dynamical instability. In different settings, the charge has different effects. For the slowly rotating black ring, the charge lowers the angular momentum and weakens the instability. While for the MP black hole, the charge also lowers the angular momentum  and may enhance the instability.  For the slowly boosted black string the charge does not affect the angular momentum but weakens the GL instability.

As we showed, the extremal black hole needs special consideration. It is a singular point in the solution space, in the sense that the extreme point at the leading order requires vanishing angular velocity. This is related to the fact that the rotation is very slow and the leading order condition on the extremality gives constraint only on the charge $1-4Q^2=0$, which suggests that there must be no rotation at all. This is in accord with  the cosmic censorship. In order to discuss the near-extremal case, one has to take the extreme limit properly. In the charged black ring case, the resulting spacetime may have a null singularity, while in the charged black hole case, the spacetime seems to be regular.

The work in this paper can be extended in several directions.  For example as discussed in \cite{BRlargeD}, for the neutral large $D$ black ring  the non-uniform black ring is  suggested
to be the endpoint of the black ring evolution due to the instability. For the charged case the picture could be similar. One can  consider the numerical evolution of the effective equations to find convincing evidence. Moreover, in \cite{ElasticLargeD}, the  elastic instability are found by adding $1/n^2$ correction to the effective equations, we believe such phenomena should occur in the charged black ring as well. It would be interesting to investigate if the charge can weakens  the elastic instability.  Another extension is to consider the black ring solution in the Lovelock theory. If the rotation is still as small as the one in the Einstein theory, then the construction of the corresponding black ring solution might be feasible. It would be interesting  to study the radiation of the black ring solution in the framework of membrane paradigm as well\cite{Bhattacharyya:2016nhn}.

\vspace*{10mm}
\noindent {\large{\bf Acknowledgments}}\\

The work was in part supported by NSFC Grant No.~11275010, No.~11335012 and No.~11325522.

\end{document}